\documentclass[aps,pra,showpacs,twoside,twocolumn,10pt,nofootinbib]{revtex4-1}
\usepackage[colorlinks=true, citecolor=red, urlcolor=blue ]{hyperref}
\usepackage{epsfig,newlfont,amssymb,amsfonts,amsmath,bm,subfigure}
\usepackage{graphicx,geometry,times}
\usepackage{palatino,mathtools,amsthm,braket,soul,enumitem,color}
\usepackage{algpseudocode}
\usepackage[normalem]{ulem}
\algrenewcommand\algorithmicindent{1.0em}

\begin{document}
 
\title{Hierarchies of localizable entanglement due to spatial distribution of local noise}
\author{Ratul Banerjee$^1$, Amit Kumar Pal$^{2,3,4}$, Aditi Sen(De)$^1$}
\affiliation{$^1$Harish-Chandra Research Institute, HBNI, Chhatnag Road, Jhunsi, Prayagraj - 211019, India \\
$^2$ Department of Physics, College of Science, Swansea University, Swansea - SA2 8PP, United Kingdom \\
$^3$ Faculty of Physics, University of Warsaw, Pasteura 5, 02-093 Warsaw, Poland\\
$^4$ Department of Physics, Indian Institute of Technology Palakkad, Palakkad, 678557, India
}

\begin{abstract}
Complete characterization of a noisy multipartite quantum state in terms of entanglement requires full knowledge of how the entanglement content in the state is affected by the spatial distribution of noise in the state. Specifically, we find that if the measurement-basis in the protocol of computing localizable entanglement and the basis of the Kraus operator representing the local noisy channel do not commute, the information regarding the noise is retained in the system even after the qubit is traced out after measurement. Using this result and the basic properties of entanglement under noise, we present a set of hierarchies that localizable entanglement over a specific subsystem in a multiqubit state can obey when local noise acts on the subparts or on all the qubits of the whole system.
%in a specific distribution in terms of presence and absence of noise in different parts of the system.  
In particular, we propose two types of hierarchies -- one tailored according to the number of  noisy unmeasured qubits, and the other one that  depends additionally  on the cardinality of the set of noisy measured qubits, leading to the classification of quantum states.
We report the percentage of states satisfying the proposed hierarchies in the case of random three- and four-qubit systems and show, using both analytical methods and numerical simulations,  that in almost all the cases,  anticipated hierarchies tend to hold with the variation of the strength of noise.   
%, and comment on the comparative behaviour of the two. Our work also results in important classification of states in the space of three- and four-qubit states in terms of these hierarchies.     
\end{abstract}

\maketitle

\section{Introduction}
\label{sec:intro}

Quantum entanglement~\cite{horodecki2009}, in both its bipartite and multipartite form, has been proved to be an important ingredient in quantum information processing tasks, including quantum teleportation~\cite{bennett1993,bouwmeester1997,murao1999,bouwmeester2000,grudka2004,sende2010a}, quantum dense coding~\cite{bennett1992,mattle1996,sende2010,bruss2004,bruss2006,das2014,das2015,horodecki2012,
shadman2012}, entanglement swapping~\cite{pan1998,schmid2009}, quantum cryptography~\cite{ekert1991,jennewein2000,cleve1999,karlsson1999,gisin2002,pirandola2019,hillery1999}, quantum metrology~\cite{giovannetti2011,toth2014,dowling2015,demkowicz2015,pezze2017}, and measurement-based quantum computation~\cite{raussendorf2001,briegel2009,hein2004,hein2006}. Along with designing photonic setups for performing quantum protocols~\cite{raimond2001,prevedel2009,gao2010,yao2012,pan2012,wang2016,wang2018},  quantum many-body systems  such as trapped ions~\cite{leibfried2003,haffner2008,singer2010,duan2010,monz2011,friis2018}, superconducting qubits~\cite{barends2014,yang2016,gong2019}, nuclear magnetic resonance molecules~\cite{vandersypen2005,negrevergne2006}, ultracold atoms in optical lattices~\cite{mandel2003,leibfried2005,bloch2012,cramer2013}, and solid-state systems~\cite{bradley2019} are also potential candidates for realizing quantum computational tasks as well as  quantum transport over a short distance. This has also led to the study of entanglement properties in the characteristic phases of paradigmatic quantum many-body systems~\cite{amico2008,chiara2018}, especially in the vicinity of quantum criticality.  However, successful experimental realizations suffer from environmenntal interactions with the system, thereby reducing the entanglement content in the system. This has motivated rigorous investigation in understanding the behaviour of entanglement when different types of noise is present in the system~\cite{zyczkowski2001,dodd2004,yu2004,almeida2007,salles2008,yu2009}. 

Recent emergence of various noisy intermediate-scale quantum (NISQ) devices~\cite{preskill2018,paler2018,nash2019} has highlighted the need for appropriately characterizing the quantum states, which are prepared in these systems, and are envisioned as resources in quantum protocols. Note that all of them are currently constituted of less than 100 qubits  and are viewed as the potential pathways to achieve “quantum supremacy”~\cite{preskill2012}.
% has highlighted the need for appropriately characterizing the quantum states, which are prepared in these systems, and are envisioned as resources in quantum protocols. 
A major step towards such characterization is the investigation of the spatial distribution of entanglement in these multipartite systems, subject to the presence of noise in different parts of the system. Spatial distribution of entanglement has recently been proven useful in  enabling Einstein-Podolsky-Rosen (EPR) steering~\cite{cavalcanti2016} in atomic clouds~\cite{kunkel2018} and in Bose-Einstein condensate~\cite{fadel2018}.  Moreover, entanglement has been generated between atoms occupying different spatial regions of a multipartite system composed of thousands of ultracold atoms~\cite{lange2018}, which puts the importance of the study of the effect of noise present in different spatial parts of the system into perspective. Apart from the many-body systems, inspiration of such studies can also be found in the possible relation between the quantum yield of a light-harvesting complex~\cite{sarovar2010,zhu2012,lambert2013,chanda2014} and the spatial distribution of entanglement among its different components~\cite{fassioli2010}. 

Despite extensive studies on the effect of local as well as global noise on the entanglement content of a multiparty system~\cite{zyczkowski2001,dodd2004,yu2004,almeida2007,salles2008,yu2009}, it is not yet clear how the spacial distribution of noise, in the form of the presence and absence of \emph{local} noise at various parts of a multiparty system, affects the entanglement content of the whole system as well as a certain block of the system, which may consist of two or more sites. While the notion of \emph{locality of noise} is well-established~\cite{nielsen2010,holevo2012}, a major issue towards the line of investigation of the latter is the quantification and subsequent computation of entanglement in different parts of a multiparty quantum system. In this paper, we focus on the maximum average entanglement that can be localized in subparts of a multiparty system by performing local projection measurement on the rest of the system, which is also referred to as the \emph{localizable entanglement} (LE)~\cite{verstraete2004,verstraete2004a,popp2005,jin2004} (cf.~\cite{divincenzo1998}). It has been shown to be an appropriate quantifier in measuring entanglement since it still  keeps information about the  quantum correlations in the  measured subparts of the  original system. In particular, the measure is made for the multiqubit  Greenberger-Horne-Zeilinger (GHZ)~\cite{greenberger1989} and stabilizer states~\cite{nielsen2010,amaro2018,amaro2019}, and states emerging in the studies of quantum networks and entanglement percolation (see ~\cite{acin2007} and references thereto). Also, the potential of localizable entanglement, even when computed over a pair of qubits, to be considered as a multipartite measure of entanglement~\cite{gour2006,banerjee2019} makes it an appropriate candidate for the investigation of the effect of spatial distribution of noise on the entanglement at different parts of the system.

In this paper, we establish the hierarchies of the values of localizable entanglement based on the sites on which local noise acts and  local measurements are performed to different subparts of a multiparty quantum system. We show that the proposed ranking is independent of the strength of the noise. 
% We show that the  ranking of LE  for different situations changes depending  on the sites on which local noise act, and  local measurements are performed. 
We divide the ranking of states into two categories. One of  them depends only on the cardinality of the set of noisy qubits where measurements are not performed, and we call it as \emph{envelope ranking}. On the other hand, there can also  be a \emph{fine-grained hierarchy}, which additionally depends on the cardinality of set of noisy qubits on which local measurements are performed. To demonstrate this classification among states, we consider local uncorrelated Pauli noise including the bit-flip, phase-flip, and depolarizing noise as the non-dissipative ones, and local amplitude-damping noise as an example of the dissipative noise. We analytically derive conditions under which the information about the local Pauli noise on the measured qubits is re-encoded in the system even when the measured subsystems are traced out in the computation of localizable entanglement.  For  simplifying  the investigation, we use the restricted localizable entanglement (RLE)~\cite{amaro2018,amaro2019} in  which  local measurements are restricted  to spin measurements based on Pauli  matrices.  When   three-qubit  states belonging to the paradigmatic generalized GHZ and W states, GHZ - and W-class are subjected to local noise, we compute the percentage of states satisfying fine-grained and envelope hierarchies for LE and RLE. 
%Using these results, we investigate the hierarchies between the values of localizable  as well as restricted localizable entanglement. 
We also discuss the existence of characteristic noise strengths in relation to the vanishing of the RLE and LE,  and point out its relation with the hierarchies. We extend our results to Haar uniformly generated random four-qubit systems, and compare the results with three-qubit random states  regarding the validity of this characterization. We also observe that rankings of LE in states  based on only the cardinality of the set of noisy qubits fail with increase of noise. 
% We also show a correspondence between the hierarchy based on the cardinality of the set of noisy qubits and the hierarchy designed specially for localizable entanglement when the strength of the local noise is increased. 

The rest of this paper is organized as follows. In Sec.~\ref{sec:def}, we provide necessary definitions of localizable and restricted localizable entanglement, and different local noise models considered in this paper. The effect of local noise on restricted localizable entanglement, when noise is applied to the whole or a group of qubits in the system, is described in Sec.~\ref{sec:effecct_of_nnoise_RLE}. The hierarchies of localizable and restricted localizable entanglement has been introduced in Sec.~\ref{sec:hierarchies}, and the validity of them in the systems of three and four qubits has been discussed in Sec.~\ref{sec:bf-pf}-\ref{sec:ad} in a case-by-case basis. Sec.~\ref{sec:conclude} contains the concluding remarks.

\section{Definitions  and Formalism}
\label{sec:def}

In this section, we briefly discuss localizable entanglement, and the issue of its optimization.  We also define terminologies used while considering different types of local noisy channels.

\subsection{Localizable entanglement}
\label{subsec:le}

In a multiqubit system constituted of $N$ qubits, the  maximum possible average entanglement that can be accumulated over a chosen set $S$ of $N-n$ qubits by performing independent local projection measurements on the rest of the $n$ qubits forming the set $R$, with $R\cup S=\emptyset$, is called the localizable entanglement (LE)~\cite{verstraete2004,verstraete2004a,popp2005,jin2004} over the qubits in $S$.  Let us denote the qubits in the $N$-qubit system by $1,2,\cdots,N$, and an arbitrary qubit by $i$, $i=1,2,\cdots,N$. Without any loss in generality, we always assume that the measurement is performed over the last $n$ qubits, such that $i\in R\equiv\{N-n+1,N-n,N-n-1,\cdots,N-1,N\}$. 
%among which the $n$ measured qubits  are designated by the indices $r_1,r_2,\cdots,r_n$, and the rest of the qubits are labelled as $s_1,s_2,\cdots,s_{N-n}$, with $s_i,r_i\in\{1,2,\cdots, N\}$. 
For a quantum state $\rho$ describing the $N$-qubit system, the LE over the set $S$ of qubits is given by 
\begin{eqnarray}
E_{S}=\max \sum_{k=0}^{2^n-1}p_kE(\tilde{\rho}^{k}_S). 
\end{eqnarray}
Here, the maximization is performed over the complete set of single-qubit  rank-$1$ projection measurements on the qubits in $R$. The multi-index $k\equiv{k_{N-n+1}k_{N-n}\cdots k_{N}}$ denotes the outcome of the measurement corresponding to the projectors $\{P_{i}^{k_i}=\ket{k_{i}}\bra{k_{i}}\}$ on the qubits $i\in R$. The reduced state  $\tilde{\rho}_S^{(k)}$  of the qubits in $S$ is obtained by tracing out the qubits in $R$ from the post-measured state $\tilde{\rho}^{k}$ corresponding to the outcome $k$, given by 
\begin{eqnarray}
\tilde{\rho}^{k}=\frac{1}{p_k}\mathcal{M}_{k}\rho\mathcal{M}_k^\dagger.
\label{eq:le}
\end{eqnarray}
The probability of obtaining the  measurement-outcome $k$ is $p_k=\text{Tr}\left[\mathcal{M}_{k}\rho\mathcal{M}_k^\dagger\right]$, and the measurement element is given by 
\begin{eqnarray}
\mathcal{M}_k=\bigotimes_{i\in R}P_i^{k_{i}}\bigotimes_{j\in S}I_{j}, 
\end{eqnarray}
with $I_{j}$ being the identity operator in the Hilbert space of qubit $j$. 

In the case of qubit systems, the rank-$1$ projectors corresponding to each qubit $i\in R$ can be parametrized using two real parameters $\theta_{i}$ $(0\leq\theta_{i}<\pi)$ and $\phi_{i}$ $(0\leq\phi_{i}\leq 2\pi)$  as $P_{i}^{k_i}=\ket{k_i}\bra{k_i}$, $k_i=\mathbf{0},\mathbf{1}$, with~\cite{nielsen2010} 
\begin{eqnarray}
\ket{\mathbf{0}}_i&=&\cos\frac{\theta_i}{2}\ket{0}_i+\text{e}^{\text{i}\phi_i}\sin\frac{\theta_i}{2}\ket{1}_i,\nonumber\\
\ket{\mathbf{1}}_i&=&\sin\frac{\theta_i}{2}\ket{0}_i-\text{e}^{\text{i}\phi_i}\cos\frac{\theta_i}{2}\ket{1}_i,
\label{eq:real_parameters}
\end{eqnarray}
where $\{\ket{0}_i,\ket{1}_i\}$ is the computational basis of the Hilbert space of qubit $i$. This parametrization reduces the maximization in Eq. (\ref{eq:le}) to a maximization problem involving $2n$ real parameters. However, the maximization becomes  challenging when $n$ is a large integer~\cite{verstraete2004,verstraete2004a,popp2005,jin2004,sadhukhan2017}. There exists only a number of systems  for which the optimal measurement basis for maximizing LE can be determined analytically, viz. a number of paradigmatic quantum states including the multi-qubit Greenberger-Horne-Zeilinger (GHZ)~\cite{greenberger1989,sadhukhan2017}, the W~\cite{zeilinger1992,dur2000,sadhukhan2017}, the Dicke~\cite{dicke1954,kumar2017,sadhukhan2017}, and the stabilizer~\cite{hein2004,hein2006,amaro2018,amaro2019} states,  and quantum spin Hamiltonians with certain symmetries~\cite{venuti2005}. 

The definition of LE (Eq.~(\ref{eq:le})) depends also on the computability of a chosen entanglement measure $E$, which is called the seed measure~\cite{sadhukhan2017}, for the reduced state $\rho_S$. In cases where one has to deal with a mixed state describing the $N$-qubit system, such as the scenarios involving noise, subsequent reduced post-measured states $\rho_S$ are also mixed. The scarcity of computable entanglement measures for mixed states in arbitrary dimension~\cite{horodecki2009} makes the determination of LE difficult in these situations. In this paper, we  restrict ourselves to the cases where $n=2$, for which several computable entanglement measures are available~\cite{horodecki2009}. We select negativity~\cite{vidal2002} as the entanglement measure for calculating LE, which, for a generic bipartite state,  $\varrho_{ab}$, describing parties $a$ and $b$, is defined as 
\begin{eqnarray}
E(\varrho_{ab})=\left|\left|\varrho_{ab}^{\text{T}_a}\right|\right|_1-1.
\label{eq:neg}
\end{eqnarray}
Here, $\left|\left|\varrho_{ab}^{\text{T}_a}\right|\right|_1$ is the trace-norm of $\varrho_{ab}^{\text{T}_a}$, which is obtained by performing a partial transposition of the state $\varrho_{ab}$ with respect to the party $a$.  It can be shown~\cite{peres1996,horodecki1996} that the negativity $E(\varrho_{ab})$ can be computed from the eigenvalues $\{\lambda_i\}$ of $\varrho_{ab}^{\text{T}_a}$ as the absolute sum of the negative eigenvalues, given by
\begin{eqnarray}
E(\varrho_{ab})=\sum_{\lambda_i<0}\left|\lambda_i\right|.
\label{eq:neg_eigen}
\end{eqnarray}
Since $\varrho_{ab}^{\text{T}_a}$ has only one negative eigenvalue when $\varrho_{ab}$ describes a two-qubit state~\cite{sanpera1998}, $E(\varrho_{ab})=|\lambda|$ for $\lambda<0$.

\subsection{Restricted localizable entanglement}
\label{subsec:rle}

In many of the systems where analytical computation of LE over a pair of qubits is possible, the optimal bases corresponding to the local projection measurements on the rest of the qubits belong to the eigenvectors of the Pauli matrices, $\sigma^{x,y,z}$~\cite{sadhukhan2017,hein2004,hein2006,amaro2018,amaro2019,venuti2005}. These results inspire the following assumption, and the subsequent definition of a \emph{restricted} LE (RLE)~\cite{amaro2018,amaro2019}, which is obtained by allowing only Pauli projections over the qubits $i\in R$ (cf. restricted quantum discord~\cite{chanda2015}).

\noindent\textbf{Assumption:} \emph{Corresponding to each of the qubits $r_j\in R$, projection measurements ``only" in the basis of (i) $\sigma^z_i$  ($\theta_i=\phi_i=0$), or (ii) $\sigma^x_i$ ($\theta_i=\frac{\pi}{2},\phi_i=0$), or (iii) $\sigma^y_i$ ($\theta_i=\phi_i=\frac{\pi}{2}$) are allowed in order to accumulate entanglement on the qubits in $S$.} 

\noindent The real parameters $\{\theta_i,\phi_i\}$ are defined in Eq.~(\ref{eq:real_parameters}), and the subsequent discussion.  Evidently, under the above assumption, there can be a total of $3^n$ combinations of Pauli bases on the $n$-qubits in $R$, for each of which $2^n$ measurement outcome is possible and an average entanglement, representing a possible value of  RLE, can be computed.

Let us now denote the Pauli matrix corresponding to the measurement bases on the qubit $i\in R$ by $\sigma_i^{\alpha_i}$, where values of $\alpha_i$, given by $\alpha_i=0,1,2$, represent the Pauli matrices $\sigma^x_i$, $\sigma_i^y$, and $\sigma_i^z$, respectively.  The overall Pauli measurement configuration over the region $R$ is represented by $\sigma^\alpha_R$, where $\alpha\equiv \alpha_{N-n+1}\alpha_{N-n}\cdots\alpha_N$ is the multi-index having values $0,1,2,\cdots,3^n-1$.  For each of the all possible measurement combinations $\{\sigma^\alpha_R;\alpha=0,1,\cdots,3^n-1\}$, one can compute a value of the RLE, denoted by $E_{\alpha,S}^\prime$. The maximum value of RLE, denoted by
\begin{eqnarray}
E_S^\prime=\underset{\{\sigma_R^\alpha\}}{\max}E_{\alpha,S}^\prime,
\label{eq:rle}
\end{eqnarray}
is obtained by maximizing $E_{\alpha,S}^\prime$ over the complete set of Pauli measurement configurations $\{\sigma_R^\alpha,\alpha=0,1,\cdots,3^n-1\}$. From the definition of LE, we have
\begin{eqnarray}
E_{\alpha,S}^\prime\leq E_S^\prime\leq E_S
\end{eqnarray}

The importance of $E_S^\prime$ lies in the existence of quantum states, such as the stabilizer states without~\cite{hein2004,hein2006} and in the presence~\cite{amaro2018,amaro2019} of local uncorrelated Pauli noise, and ground states of certain quantum many-body systems~\cite{verstraete2004,verstraete2004a,popp2005,jin2004,sadhukhan2017,venuti2005}, for which $E_S^\prime=E_S$. Moreover, if one now considers the absolute error originated due to the restriction, given by $|E_S-E^\prime_S|$, with $|E_S-E^\prime_S|\leq \varepsilon$, $\varepsilon$ being a small number, typically $\sim 10^{-3}$ or less, then the LE can be safely approximated by the RLE. In this situation, the definition of RLE can be used to obtain closed form expressions, which represents the LE with negligible error, and which can not be obtained analytically otherwise. This will be clear in subsequent sections.

\subsection{Models of uncorrelated noise}
\label{subsec:uncorrelated}

We shall focus on local noise models in this paper, where the noise is confined at and is identical for individual qubits of the total system. We assume a scenario where single-qubit uncorrelated noise acts on $m$ $(m\leq N)$ qubits 
%denoted by $l_1,l_2,\cdots,l_m$ and forming the set $L=\{l_1,l_2,\cdots,l_m\}$, 
in the $N$-qubit system, forming the set $L$. For a fixed value of $m$, there can be $\binom{N}{m}$ such noise configurations with $m=0,1,2,\cdots,N$ (see Fig. \ref{fig:4qubit_h13} for an example of a four-qubit system). We shall show  the interplay between the set of qubits, $R$, on which the measurements are made,  and the set, $L$, on which the noise acts.  Let us denote an $N$-qubit quantum state by $\rho_N^m$, where the subscript and the superscript specify the number of qubits in the system and the number of noisy qubits respectively. The noiseless state is represented by $\rho_N^0$ in this notation. The noise map, for the initial $N$-qubit state $\rho_N^m$, is given by 
\begin{eqnarray}
\rho_N^0\rightarrow\rho_N^m=\Lambda_L(\rho_N^m).
\label{eq:evolve_0}
\end{eqnarray}

We assume \emph{uncorrelated} single-qubit noisy channels, and employ the Kraus operator representation for the evolution $\Lambda_{L}$ of a multiqubit state $\rho_N^0$, where the operation $\Lambda_{L}(.)$ can be expressed by an operator-sum decomposition given by~\cite{nielsen2010,holevo2012} 
\begin{eqnarray}
 \rho_N^m&=&\sum_{\mu=0}^{d^m-1}\left(I_{N-m}\otimes K_\mu\right)\rho_N^0\left(I_{N-m}\otimes K^\dagger_\mu\right),\nonumber \\ 
\label{eq:evolve}
\end{eqnarray}
with $\{K_\mu=\sqrt{p_\mu}\tilde{K}_\mu\}$ being the Kraus operators satisfying $\sum_{\mu}K_\mu^\dagger K_\mu=I$, and 
\begin{eqnarray}
\tilde{K}_\mu=\bigotimes_{i\in L} K_{\mu_i},\;p_\mu=\prod_{i\in L} p_{\mu_i},
\label{eq:kraus_pauli}
\end{eqnarray}
where $\sum_{\mu_i=0}^{d-1}p_{\mu_i}=1$ for a specific $i\in L$, and $\mu\equiv \cdots \mu_{i-1}\mu_i\mu_{i+1}\cdots$ is the multi-index corresponding to the $m$-qubit Kraus operators for the qubits $i\in L$. Here, $I_{N-m}=\bigotimes_{i\notin L}I_i$ is the identity operator in the Hilbert space of the subsystem of $(N-m)$ noiseless qubits, $\{K_{\mu_i};\mu_i=0,1,\cdots,d-1\}$ is the set of Kraus operators corresponding to the noisy channel on the qubit $i$, and $d$ is the cardinality of the set $\{K_{\mu_i}\}$. In this paper, we shall focus on non-dissipative  single-qubit Pauli noise including the bit-flip (BF), phase-flip (PF), bit-phase-flip (BPF), and depolarizing (DP) channels~\cite{nielsen2010,holevo2012}, while the amplitude-damping (AD) channel~\cite{nielsen2010,holevo2012} is considered as an example of a dissipative noise.  The single-qubit Kraus operators corresponding to these channels for an arbitrary qubit $l_i$ are given by 
\begin{widetext}
\begin{eqnarray}
\text{BF Channel: } & d=2; & K_{0}=\sqrt{1-\frac{p}{2}}I_i,\, K_{1}=\sqrt{\frac{p}{2}}\sigma^x_i;\nonumber \\ 
%\text{BPF Channel: } & d=2; & K_{\mu_{l_i}=0}=\sqrt{1-\frac{p}{2}}I_{l_i},\, K_{\mu_{l_i}=1}=\sqrt{\frac{p}{2}}\sigma^y_{l_i};\nonumber \\ 
\text{PF Channel: } & d=2; & K_{0}=\sqrt{1-\frac{p}{2}}I_i,\, K_{1}=\sqrt{\frac{p}{2}}\sigma^z_i;\nonumber \\ 
\text{DP Channel: } & d=4; & K_{0}=\sqrt{1-\frac{3p}{4}}I_i,\, K_{1}=\sqrt{\frac{p}{4}}\sigma^x_i,\, K_{2}=\sqrt{\frac{p}{4}}\sigma^y_i,\, K_{3}=\sqrt{\frac{p}{4}}\sigma^z_i;\nonumber \\ 
\text{AD Channel: } & d=2; & K_{0}=\left(\begin{array}{cc}
   1 & 0 \\
   0 &  \sqrt{1-p}\\
  \end{array}\right),
K_{1}=\left(\begin{array}{cc}
   0 & \sqrt{p} \\
   0 &  0\\
  \end{array}\right),
\label{eq:channel_prob}
\end{eqnarray}
\end{widetext}
where the subscripts of the Kraus operators $K$ are the different values of $\mu_i$, $I_i$ is the identity matrix in the Hilbert space of qubit $i$, and  $p$ ($0\leq p\leq 1$) can be interpreted as the strength of the noise.

\section{Effect of local Pauli noise on restricted localizable entanglement}
\label{sec:effecct_of_nnoise_RLE}

In this section, we shall discuss the effect of local Pauli noise on the restricted localizable entanglement (see Sec.~\ref{subsec:rle}) of an arbitrary noisy quantum state $\rho_N^m$. Later, we shall show that there exists quantum states for which these results can safely describe the same for localizable entanglement with negligible error. 

Computation of RLE in quantum states subjected to local Pauli noise requires a projection measurement in the basis of a chosen Pauli matrix $\sigma^{\alpha_j}_j$ on a qubit $j$ in the noisy state $\rho_N^m$. This measurement is followed by a partial trace operation on the same qubit.  For demonstration, we choose the BF noise, where the noisy state $\rho_N^m$, obtained from the noiseless $N$-qubit state $\rho_N^0$ by the application of the BF noise, can be written as 
%\begin{widetext}
\begin{eqnarray}
\rho_N^m&=&\left(1-\frac{p}{2}\right)^m\rho_N^0+\left(1-\frac{p}{2}\right)^{m-1}\frac{p}{2}\sum_{\forall i\in L}\sigma_i^{x}\rho_N^0\sigma_i^{x}\nonumber\\
&&+\left(1-\frac{p}{2}\right)^{m-2}\left(\frac{p}{2}\right)^{2}\sum_{\underset{i\neq j}{\forall i,j\in L}}\sigma_i^{x}\sigma_j^x\rho_N^0\sigma_i^{x}\sigma_j^x\nonumber \\ &&+\cdots\nonumber  \\
&& +\left(\frac{p}{2}\right)^{m}\left[\bigotimes_{\forall i\in L}\sigma^x_i\right]\rho_N^0\left[\bigotimes_{\forall i\in L}\sigma^x_i\right].
\label{eq:pauli_noisy_bf}
\end{eqnarray}
%\end{widetext}
We assume a projection measurement on the qubit $j$ in the basis of a chosen Pauli operator $\sigma^{\alpha_j}_j$, where the index $\alpha_j$ has been defined in the discussion preceding Eq.~(\ref{eq:rle}). The projection operation can be written as 
\begin{eqnarray}
P^{\alpha_j}_{k_j}=\frac{1}{2}\left[I_j+(-1)^{k_j}\sigma^{\alpha_j}_j\right],
\label{eq:pauli_projection_prop}
\end{eqnarray} 
where $k_j=0,1$ represents the measurement outcomes corresponding to the bases of $\sigma^{\alpha_j}$. From the properties of Pauli operators,  
\begin{eqnarray}
\sigma^{\gamma}_jP_{k_j}^{\alpha_j}\sigma^{\gamma}_j=P_{k_j^\prime}^{\alpha_j},
\label{eq:projection_identity}
\end{eqnarray}
with $\gamma=0,1,2$, where $k^\prime_j=k_j$ if $\gamma=\alpha_j$, and $k^\prime_j=k_j+1$ modulo $2$ if $\gamma\neq\alpha_j$. Note that $\gamma=\alpha_j$ describes the situation where the projection operator $P^{\alpha_j}_{k_j}$ and $\sigma^\gamma_j$ have the same basis, while  $\gamma\neq\alpha_j$ indicates otherwise. While the projection measurement destroys all quantum correlation between the qubit and the rest of the system, the information regarding the noise on the measured qubit depends both on the basis of the projection measurement as well as the basis of the single-qubit Pauli noise. We formulate this via the following proposition.

\noindent
\begin{minipage}{8.5cm}

\noindent\textbf{Proposition I.} \emph{When a projection operation $P^{\alpha_j}_{k_l}$ in the basis of a chosen Pauli matrix $\sigma^{\alpha_l}$, $\alpha_l=0,1,2$,  is performed on a chosen qubit $l$ corresponding to an $N$-qubit state $\rho_N^m$ obtained by the application of local uncorrelated  bit-flip noise on $m$ qubits forming a set $L$, followed by a tracing out of the measured qubit $l$, then for $l\in L$, the information regarding the noise on the measured qubit encoded in the probabilities corresponding to the Kraus operators  is lost if  the Pauli operator chosen for measurement matches with Kraus operator corresponding to the local bit-flip noise which is not identity, i.e., if $\alpha_{l}=0$, and is retained in the rest of the system otherwise, i.e.,  if $\alpha_l=1,2$.} 

\end{minipage} 

\noindent\emph{Proof.}  First, we note that two situations are possible corresponding to the chosen qubit $l$.   First, we assume that $l\notin L$, i.e., the measured qubit is noiseless. This is a situation where application of the projection operation on qubit $l$ and subsequent tracing out of the measured qubit results in an $(N-1)$-qubit post-measured states on the rest of the qubits, with all the noisy qubits in $L$  present in the system. The post-measured states corresponding to both the outcomes $k_l=0,1$ on the remaining $N-1$ qubits after tracing out the qubit $l$ are of the form given in Eq.~(\ref{eq:pauli_noisy_bf}), where the information regarding the noise on qubit $l$ is lost.  

As the second situation, we consider the case $l\in L$, i.e., the situation described in the proposition.   
%and without any loss in generality, assume that $r_j=l_1$. 
In this scenario, two possibilities exist.

\begin{center}\underline{\emph{Case 1.}  $\gamma=\alpha_l$.} \end{center}

\noindent This situation occurs in the case of BF noise when $\alpha_l=0$. Application of $P^0_{k_l}$ on $\rho_N^m$ leads to the post-measured states on the remaining $N-1$ qubits as 
%\begin{widetext} 
\begin{eqnarray}
\tilde{\rho}_{N-1}^{m-1,k_l}&=&\left(1-\frac{p}{2}\right)^{m-1}\rho_{N-1}^{0,k_{l}}\nonumber \\ 
&&+\left(1-\frac{p}{2}\right)^{m-2}\frac{p}{2}\sum_{\forall i\in L\backslash l}\sigma_{i}^{x}\rho_{N-1}^{0,k_{l}}\sigma_i^{x}\nonumber \\
&&+\left(1-\frac{p}{2}\right)^{m-3}\left(\frac{p}{2}\right)^{2}\sum_{\underset{i\neq j}{\forall i,j\in L\backslash l}}\sigma_i^{x}\sigma_j^x\rho_{N-1}^{0,k_l}\sigma_i^{x}\sigma_j^x\nonumber \\ &&+\cdots\nonumber  \\
&& +\left(\frac{p}{2}\right)^{m-1}\left[\bigotimes_{\forall i\in L\backslash l}\sigma^x_i\right]\rho_{N-1}^{0,k_l}\left[\bigotimes_{\forall i \in L\backslash l}\sigma^x_{i}\right],
\label{eq:pauli_noisy_bf_1}
\end{eqnarray}
%\end{widetext} 
for $k_{l}=0,1$, where $L\backslash l$ represents the set of noisy qubits with the qubit $l$ removed, and 
\begin{eqnarray}
\rho_{N-1}^{0,k_{l}}=\text{Tr}_{l}\left[P^0_{k_{l}}\rho_N^0 P^0_{k_{l}}\right].
\end{eqnarray} 
We remind ourselves that the superscript  ``$0$" in $\rho_{N-1}^{0,k_{l}}$ or in any other quantum state represents the fact that the state is noiseless ($m=0$), while the superscript ``$0$" in $P^0_{k_l}$ stands for $\alpha_l=0$, which implies the basis of the projection operator $P^0_{k_l}$ to be that of $\sigma^x$.  The state in Eq.~(\ref{eq:pauli_noisy_bf_1}) has a form identical to the state in Eq.~(\ref{eq:pauli_noisy_bf}). Note that this is identical to the situation where the qubit $l$ is noiseless, and the number of noisy qubits is $m-1$ in a quantum state of $N-1$ qubits.

\begin{center}\underline{\emph{Case 2.} $\gamma=\alpha_j$.} \end{center} 

\noindent This, for the BF noise, describes the case  $\alpha_l=1,2$. Eq.~(\ref{eq:projection_identity}) indicates that for half of the terms in $\rho_N^m$ (Eq.~(\ref{eq:pauli_noisy_bf})), projection operation $P^{1,2}_{k_{l}}$ leads to application of  $P^{1,2}_{k_{l}^\prime}$ on $\rho_N^0$, where $k_{l}^\prime=k_{l}+1$ modulo 2. For the rest of the terms, $P^{1,2}_{k_{l}}$ applies to $\rho_N^0$.  This results in the post-measured states of the form
\begin{eqnarray}
\tilde{\rho}_{N-1}^{m-1,k_l}&=& \left(1-\frac{p}{2}\right)\varrho^{m-1,k_l}_{N-1}+\frac{p}{2}\varrho^{m-1,k_l^\prime}_{N-1},
\end{eqnarray}
on the $N-1$ unmeasured qubits including the remaining $m-1$ noisy qubits forming the set $L\backslash l$, where $\varrho^{k_{l}}$, $k_{l}=0,1$, has the form given in Eq.~(\ref{eq:pauli_noisy_bf_1}), and $k_l^\prime=k_l+1$ modulo $2$.   Note that $\tilde{\rho}_{N-1}^{m-1,k_{l}}$ has contribution of both $\varrho^{m-1,k_{l}}_{N-1}$ (with probability $1-p/2$, which is the same as the probability with which the state of qubit $l$ under BF noise is kept unchanged) as well as $\varrho^{m-1,k_{l}^\prime}_{N-1}$ (with the same probability $p/2$ by which the state of qubit $l$ is flipped), which results from the re-encoding of the information about the single-qubit noise on qubit $l$ in the rest of the system after tracing qubit $l$ out. Hence the proof. \hfill $\blacksquare$

In situations where the noise and the projection measurement on a group of qubits, say $\mathbf{r}\equiv\{r_1,r_2,\cdots,r_{m^\prime}\}$ where $\mathbf{r}\subseteq R$ and $\mathbf{r}\subseteq L$ with $m^\prime\leq m$, have the same basis, the following corollary follows directly from \textbf{Proposition I}.

\noindent\textbf{Corollary I.1.} \emph{For a multi-qubit state as given in Eq.~(\ref{eq:evolve_0}) where $\Lambda_{L}$ represents uncorrelated identical single-qubit Pauli noise on $m$ qubits in $L$, the restricted localizable entanglement $E^\prime_{(\alpha,S)}$ where the values of $\alpha$ correspond to projection measurement on the $m^\prime$ noisy qubits in the basis that is identical with the basis of the noise, obeys the relation}
\begin{eqnarray}
E^\prime_{(\alpha,12)}(\rho_N^m)=E^\prime_{(\alpha,12)}\left(\rho_N^{m-m^\prime}\right).
\label{eq:intuitive_equality_rle}
\end{eqnarray}

\noindent\textbf{Proposition I} can be extended to the case of local projection measurements in the Pauli basis on a group of qubits in $R$. The next Proposition is for  the PF channel, having a proof similar to  that of \textbf{Proposition I}.

\noindent\textbf{Proposition II.} \emph{When a projection operation $P^{\alpha_l}_{k_l}$ in the basis of a chosen Pauli matrix $\sigma^{\alpha_l}$, $\alpha_l=0,1,2$,  is performed on a chosen qubit $l$ in a  state $\rho_N^m$ originating from local uncorrelated  phase-flip noise on $m$ qubits, and subsequently the qubit $l$ is traced out, the information regarding the noise on the measured qubit encoded in the probabilities corresponding to the Kraus operators  is lost if $\alpha_{l}=2$, and is retained in the rest of the system if $\alpha_l=0,1$, when $l$ is a noisy qubit.}

\noindent The situation, however,  is slightly different in the case of DP noise, which is given in \textbf{Proposition III}, and which can clearly be seen from the form of the corresponding Kraus operators in Eq.~(\ref{eq:channel_prob}).

\noindent\textbf{Proposition III.} \emph{When a projection operation $P^{\alpha_l}_{k_l}$ in the basis of a chosen Pauli matrix $\sigma^{\alpha_l}$, $\alpha_l=0,1,2$,  is performed on a noisy qubit $l$ in a state $\rho_N^m$ having local uncorrelated  depolarizing noise on $m$ qubits,  and subsequently a tracing out of qubit $l$ is performed, the information regarding the noise on the measured qubit encoded in the probabilities corresponding to the Kraus operators remains in the rest of the system irrespective of the values of $\alpha_l$.}

\noindent\emph{Proof.}  We proceed in a fashion similar to the proof of \textbf{Proposition I}, identifying  two possible situations  (i) $l\notin L$, and (ii) $l\in L$.  The outcome of the situation (i) is complete  loss of information, as shown in   \textbf{Proposition I}. On the other hand, in situation (ii), as before, two possibilities exist: (a) $\gamma=\alpha_l$, and (b) $\gamma=\alpha_j$.  However, in  the case of DP noise, the situations (a) or (b)  never exclusively  arise as the Kraus operators involve all  three components of the Pauli matrices. While $\alpha_l=\gamma$ for a specific value of $\gamma$,   $\alpha_l\neq\gamma$ for the rest of the values of $\gamma$.  Therefore, the  information regarding the noise on the measured qubit encoded in the probabilities corresponding to the Kraus operators remains in the rest of the system irrespective of the values of $\alpha_l$.   \hfill $\blacksquare$

%\noindent For a demonstration of \textbf{Propositions I-III} in the case of a three-qubit system, see Appendix~\ref{app:proj}.   
\noindent We point out here that the possible sustainability of the effect of local Pauli noise, after performing the local projection measurement and the subsequent tracing out operation, is in contrast with the complete disappearance of the effect of noise when the noisy qubit is traced out without performing any measurement. The latter is guaranteed by the trace-preserving properties of the Kraus  operators used to characterize the local noise on individual qubits  (see Eqs.~(\ref{eq:channel_prob})).

\section{Setting the stage: Hierarchies of localizable entanglement}
\label{sec:hierarchies}

In this section, we discuss the possible hierarchies of the values of LE and RLE depending on the number of qubits on which local noise is applied. From now onward, unless otherwise stated, we localize entanglement over a region constituted of two specific qubits, say, $1$ and $2$, by performing local projection measurement on the rest of the $N-2$ qubits, indexed as $3,4,\cdots,N$, and forming the set $R$. To keep the notations uncluttered, we discard the subscript `$S$', and denote the LE (RLE) by $E_{12}(\rho_N^m)$ $\left(E_{12}^\prime(\rho_N^m)\right)$ in the following, where $\rho_N^m$ is the noisy state with local noise applied to $m$ of the $N$ qubits, forming the set $L$ of noisy qubits. In terms of the pair of qubits on which entanglement is localized, three different situations exist: when (i) none, (ii) any one, or (iii) both of the qubits $1$ and $2$ belong(s) to the set $L$, i.e., is (are) influenced by the local noise. These scenarios, along with the general intutions gathered about the trends of entanglerment measures under local decoherence~\cite{zyczkowski2001,dodd2004,yu2004,almeida2007,salles2008,yu2009}, motivate us to propose certain intuitive orderings amongst the values of LEs and RLEs, independent of the local noise models. In succeeding sections, for specific noise models, we shall illustrate whether the LE and the RLE follow such classifications. We assume $m$ to be the maximum cardinality of $L$, following the notations used in Sec.~\ref{sec:effecct_of_nnoise_RLE}. 
\begin{enumerate}
\item[(i)] \textbf{Scenario 1.} Let $L\subseteq R$, such that $m\leq N-2$,  i.e., qubits $1$  and $2$ are not affected by noise. There can be  $\binom{N-2}{m}$ possible sets $L$ of noisy qubits in $R$, which, in general, will correspond to different values of the LE and the RLE.  We expect the following relation  between the values of the  LE and the  RLE, as well as the cardinality of the set of noisy qubits: 
\begin{eqnarray}
\label{eq:(i)}
\max\{\mathcal{E}_{12(i)}(\rho_N^m)\} &\leq & \min\{\mathcal{E}_{12(i)}(\rho_{N}^{m^\prime})\},
\end{eqnarray}
where $\mathcal{E}=E(E^\prime)$ representing the LE (RLE), $m^\prime$ is the maximum cardinality of a different set $L^\prime$ of noisy qubits obeying the situation (i) (marked in the subscript), and we have assumed $m^\prime\leq m$ without any loss in generality. Possible scenarios with four qubits are exhibited in Fig.~\ref{fig:4qubit_h13}(a). Note that  for clarity, we denote  the  state $\rho^m_N$ by  $\rho_{l_1l_2\cdots l_m}$ for all illustrations, where the subscripts denote the qubits subjected to noise. 

\begin{figure}
\includegraphics[scale=0.5]{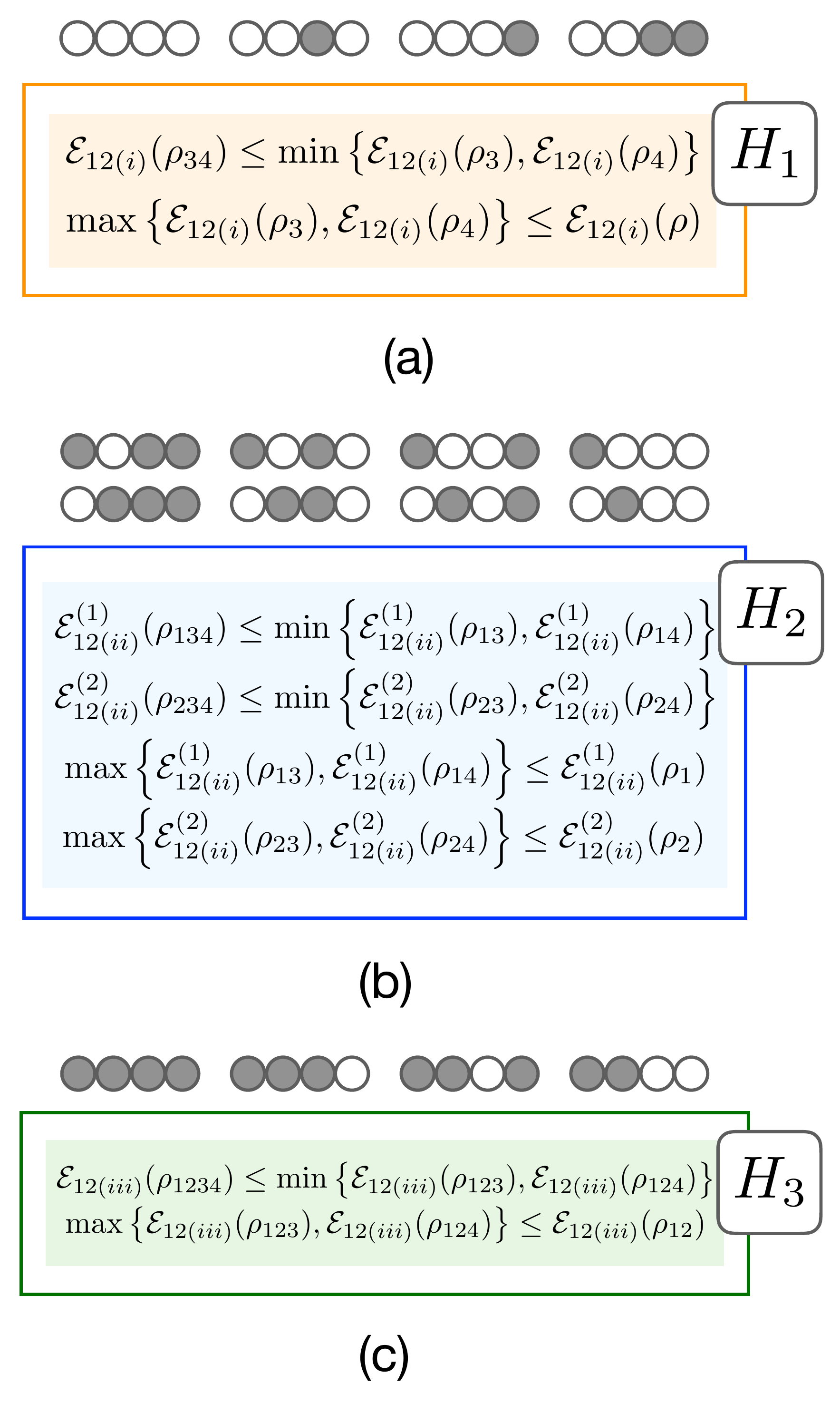}
\caption{(Colour online.) \textbf{Fine-grained hierarchies.} The hierarchies $H_1$, $H_2$, and $H_3$ for different distributions of local noise on a system of $N=4$ qubits. The configurations of noise are shown by the cluster of bubbles, where a clear (opaque) bubble represents a noiseless (noisy) qubit. The state $\rho_{l_1l_2\cdots l_m}$ implies that the qubits $\{l_1,l_2,\cdots,l_m\}$ are subjected to noise in the $N$-qubit system.}
\label{fig:4qubit_h13}
\end{figure}

\item[(ii)] \textbf{Scenario 2.} In this case, any one of the qubits $1$ and $2$ are considered to be noisy, i.e., either $1$ or $2\in L$, and consequently $m\leq N-1$. For each of the qubits $1$ and $2$ in $L$, the number of possible values of both LE and RLE for $\rho_N^m$ is $m\binom{N-1}{m-1}$. In situation (ii), we predict 
\begin{eqnarray}
\label{eq:(ii)}
\max\{\mathcal{E}_{12(ii)}^{(j)}(\rho_N^m)\} &\leq & \min\{\mathcal{E}_{12(ii)}^{(j)}(\rho_{N}^{m^\prime})\}, 
\end{eqnarray}
where the  superscript $j=1,2$ denotes the choice for noisy qubits from the unmeasured set of qubits, and $\mathcal{E}$ and $m^\prime (m^\prime\leq m)$ have similar definition as in situation (i), as depicted in Fig.~\ref{fig:4qubit_h13}(b). 

\begin{figure*}
\includegraphics[width=\textwidth]{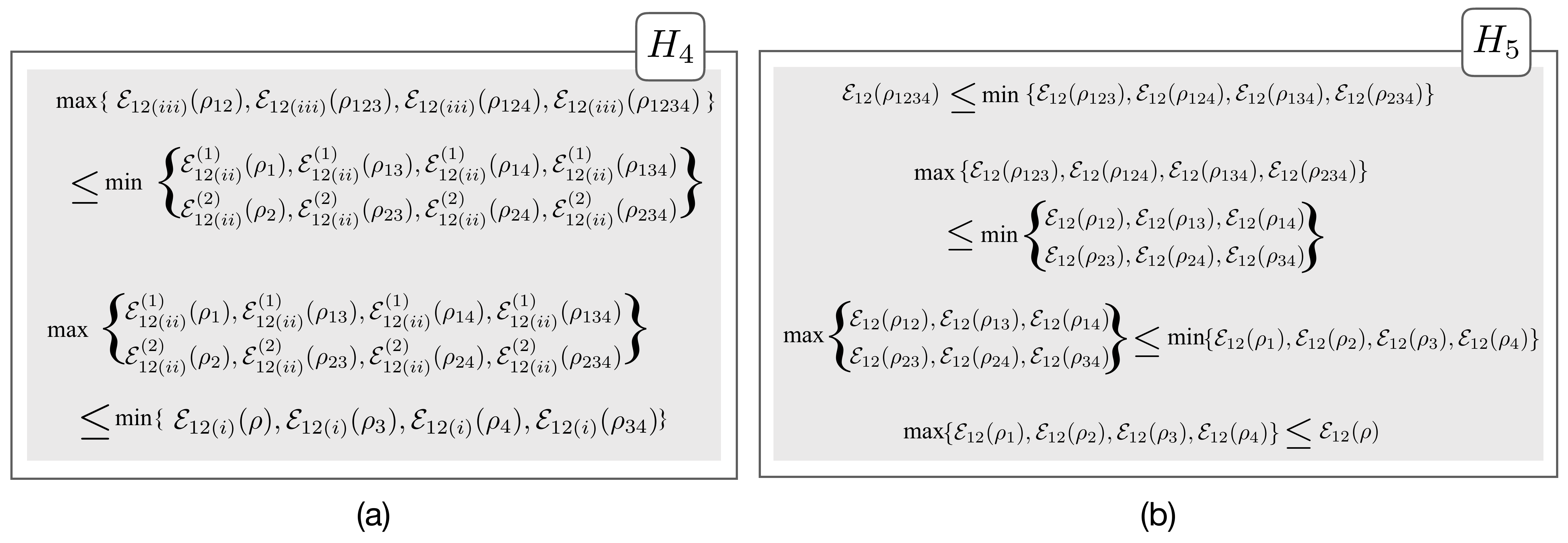}
\caption{(Colour online.) \textbf{Envelop hierarchies.} The hierarchies $H_4$ and $H_5$ for different distributions of local noise on a system of $N=4$ qubits. The interpretation of the notation $\rho_{l_1l_2\cdots l_m}$ is similar to that given in Fig.~\ref{fig:4qubit_h13}.}
\label{fig:4qubit_h45}
\end{figure*}

\item[(iii)] \textbf{Scenario 3.} Both the qubits, $1$ and $2$,  are noisy, implying $1,2\in L$, and $m\leq N$. The total number of possibilities for choosing the set of noisy qubits  $L$  from the $N$ qubits is $\binom{N-2}{m-2}$.  In this case, we anticipate 
\begin{eqnarray}
\label{eq:(iii)}
\max\{\mathcal{E}_{12(iii)}(\rho_N^m)\} &\leq & \min\{\mathcal{E}_{12(iii)}(\rho_{N}^{m^\prime})\}, 
\end{eqnarray}
where $\mathcal{E}$ and $m^\prime$ have similar definition as in situation (i), and $m^\prime\leq m$. See Fig.~\ref{fig:4qubit_h13}(c).
\end{enumerate}
\noindent And finally, between the different situations, we propose  
\begin{eqnarray}
\label{eq:le32nf}
\max\{\mathcal{E}_{12(iii)}(\rho_N^m)\} &\leq & \min\{\mathcal{E}_{12(ii)}^{(j)}(\rho_N^{m^\prime})\},\\
\label{eq:le21nf}
\max\{\mathcal{E}_{12(ii)}^{(j)}(\rho_{L})\}&\leq & \min\{\mathcal{E}_{12(i)}(\rho_N^{m^\prime})\}.
\end{eqnarray}

Note here that the inequalities (\ref{eq:le32nf})-(\ref{eq:le21nf}) together can be satisfied even when some, or none of the inequalities (\ref{eq:(i)})-(\ref{eq:(iii)}) are valid. In this sense, (\ref{eq:le32nf})-(\ref{eq:le21nf}) are considered as an \emph{envelope} over the \emph{fine-grained} hierarchies of LE presented in (\ref{eq:(i)})-(\ref{eq:(iii)}). For the purpose of comparison, as shown in Fig.~\ref{fig:4qubit_h13}, we denote inequalities (\ref{eq:(i)})-(\ref{eq:(iii)}) by $H_1$, $H_2$, and $H_3$ respectively, while the envelope inequalities (\ref{eq:le32nf})-(\ref{eq:le21nf}) together are denoted by $H_4$ (Fig.~\ref{fig:4qubit_h45}(a)). The inequalities (\ref{eq:(i)})-(\ref{eq:le21nf}) also imply that more the influence of noise on the unmeasured qubits, more can be the effect of noise on LE in the form of a reduction in its value.

We point out here that the inequalities (\ref{eq:(i)})-(\ref{eq:le21nf}) are designed with specifically LE in mind as the measure for entanglement. For a bipartite or multipartite entanglement measure other than LE~\cite{horodecki2009}, which is usually computed by using the density matrix of the whole system or the reduced density matrix of a subsystem, a more logical expectation would be a hierarchy in terms of the cardinality of the set of noisy qubits. In the present case, the ranking for entanglement over the subsystem constituted by qubits $1$ and $2$, with $m$ as the parameter, is expected to be 
\begin{eqnarray}
\max\{\mathcal{E}_{12}(\rho_N^m)\}\leq\min\{\mathcal{E}_{12}(\rho_N^{m^\prime})\},
\label{eq:old_hier}
\end{eqnarray}  
where $m\geq m^\prime$. Here, we have considered entanglement over the same pair of qubits as in the cases of (\ref{eq:(i)})-(\ref{eq:le21nf}) for the purpose of comparison. We denote (\ref{eq:old_hier}) by $H_5$. Note that $H_5$ does not take into account the configuration of noise on the qubits $1$ and $2$. However, computation of the reduced state on qubits $1$ and $2$ ensures the complete loss of information about the \emph{local} noise applied on the rest of the qubits, and the effect of noise on $\mathcal{E}_{12}$ will again be determined by the local noise present on qubits $1$ and $2$ only (see discussions succeeding \textbf{Proposition III}). The difference between this approach with the one discussed in $H_4$ is the possibility of contribution to the local noise on qubits $1$ and $2$ from the local noise on the rest of the qubits due to the projection measurement operation involved in the case of $H_4$, which is absent in $H_5$, as illustrated in Fig.~\ref{fig:4qubit_h45}. Specifically, when a large number of measured qubits are noisy, the additional contribution to noise accumulated on qubits $1$ and $2$ due to the measurement on the noisy qubits other than $(1,2)$ may be large enough so that $H_4$ and $H_5$ differs substantially (see Sec.~\ref{sec:effecct_of_nnoise_RLE} for the contribution from the measured noisy qubits).

\section{Classifications of  states with phase- and bit-flip noise}
\label{sec:bf-pf}

In this section, by considering a multi-qubit system under noise models, let us determine the hierarchies between the values of LE corresponding to different configurations of noise on the system. We remind ourselves that we have adopted a notation where the noisy  state $\rho_N^m$ is denoted by $\rho_{l_1l_2\cdots l_m}$, where the subscripts provide the positions of the noisy qubits. We start the discussion with a three-qubit system, and adopt the notation used in Figs.~\ref{fig:4qubit_h13}-\ref{fig:4qubit_h45} to describe the noisy states for clarity. The hierarchies discussed in Eqs.~(\ref{eq:le32nf})-(\ref{eq:le21nf}) in the case of a three-qubit system becomes 
\begin{eqnarray}
\label{eq:3hier_1}
&&\max\left\{\mathcal{E}_{12}\left(\rho_{123}\right),\mathcal{E}_{12}\left(\rho_{12}\right)\right\}\nonumber \\ 
& \leq & \min\left\{\mathcal{E}_{12}\left(\rho_{13}\right),\mathcal{E}_{12}\left(\rho_{23}\right),\mathcal{E}_{12}\left(\rho_{1}\right),\mathcal{E}_{12}\left(\rho_{2}\right)\right\}, \\
\label{eq:3hier_2}
&&\max\left\{\mathcal{E}_{12}\left(\rho_{13}\right),\mathcal{E}_{12}\left(\rho_{23}\right),\mathcal{E}_{12}\left(\rho_{1}\right),\mathcal{E}_{12}\left(\rho_{2}\right)\right\} \nonumber \\
&\leq &\mathcal{E}_{12}\left(\rho_3\right),
\end{eqnarray} 
while the ones in Eqs.~(\ref{eq:(i)})-(\ref{eq:(iii)}) becomes 
\begin{eqnarray}
\label{eq:3hier_3}
\mathcal{E}_{12}\left(\rho_{123}\right)&\leq & \mathcal{E}_{12}\left(\rho_{12}\right),\\ 
\label{eq:3hier_4}
\max\left\{\mathcal{E}_{12}\left(\rho_{13}\right),\mathcal{E}_{12}\left(\rho_{23}\right)\right\} & \leq & \min\left\{\mathcal{E}_{12}\left(\rho_{1}\right),\mathcal{E}_{12}\left(\rho_{2}\right)\right\},\nonumber \\ 
\end{eqnarray} 
where $\mathcal{E}=E(E^\prime)$ represents the LE (RLE). On the other hand, in terms of the cardinality of the set of noisy qubits, one should expect 
\begin{eqnarray}
\label{eq:oldh_1}
&& \mathcal{E}_{12}(\rho_{123})
\leq  \min\{\mathcal{E}_{12}(\rho_{12}),\mathcal{E}_{12}(\rho_{13}),\mathcal{E}_{12}(\rho_{23})\},\\
\label{eq:oldh_2}
&&\max\{\mathcal{E}_{12}(\rho_{12}),\mathcal{E}_{12}(\rho_{13}),\mathcal{E}_{12}(\rho_{23})\}\nonumber \\
&\leq & \min\{\mathcal{E}_{12}(\rho_{1}),\mathcal{E}_{12}(\rho_{2}),\mathcal{E}_{12}(\rho_{3})\},\\
 \label{eq:oldh_3}
&&\max\{\mathcal{E}_{12}(\rho_{1}),\mathcal{E}_{12}(\rho_{2}),\mathcal{E}_{12}(\rho_{3})\} 
\leq  \mathcal{E}_{12}(\rho). 
\end{eqnarray}  
according to Eq.~(\ref{eq:old_hier}). For future references, we denote Eqs.~(\ref{eq:3hier_1})-(\ref{eq:3hier_2}) together by ``Env", and Eqs.~(\ref{eq:3hier_3}) and (\ref{eq:3hier_4}) by ``A" and ``B" respectively, while Eqs.~(\ref{eq:oldh_1})-(\ref{eq:oldh_3}) together are represented by ``C". We shall now prove whether such inequalities hold for a class of three-qubit states.

\begin{figure*}
\includegraphics[width=0.7\textwidth]{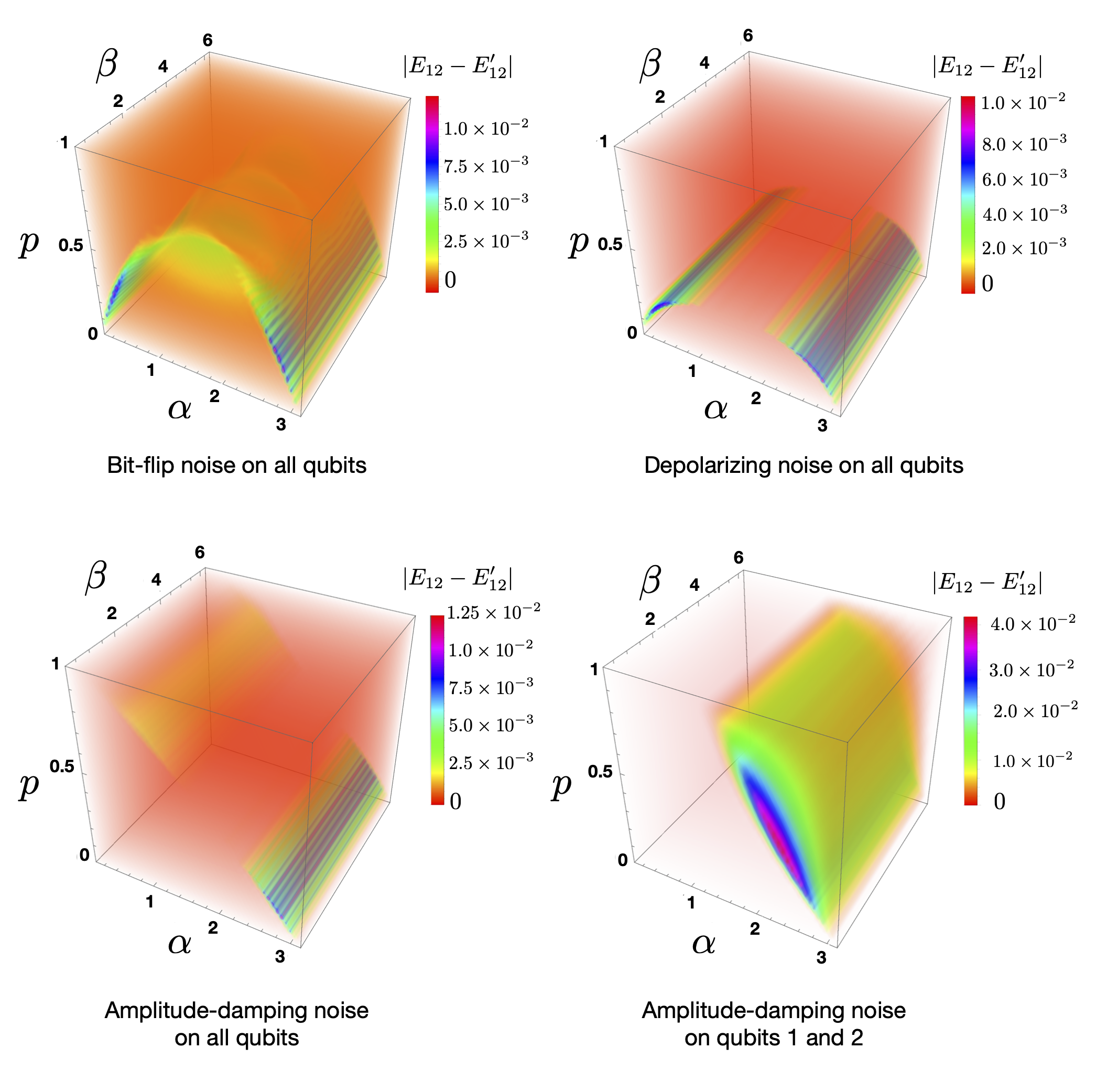}
\caption{(Colour online.) \textbf{Absolute errors for gGHZ state under noise.} The absolute value of the difference between the RLE and the LE,  $\left|E_{12}-E^\prime_{12}\right|$, as a function of $p$, $\alpha$, and $\beta$ when single qubit bit-flip, depolarizing, and amplitude-damping channels are applied to all, or a subset of the three qubits constituting a gGHZ state. All quantities plotted are dimensionless, except $\alpha$ and $\beta$, which are in radians.}
\label{fig:errors}
\end{figure*}

\noindent\textbf{gGHZ states.} Let  us first consider a paradigmatic class of three-qubit states, namely, the generalized GHZ (gGHZ) state,  given by 
\begin{eqnarray}
\ket{\text{gGHZ}}&=& \cos\frac{\alpha}{2}\ket{000}+\text{e}^{\text{i}\beta}\sin\frac{\alpha}{2}\ket{111},
\label{eq:gGHZ}
\end{eqnarray}
where $\alpha$ ($0\leq \alpha\leq \pi$) and $\beta$ ($0\leq \beta\leq 2\pi$) are real numbers. The three-qubit GHZ state is a special case of the gGHZ state with $\beta=0$, $\alpha=\frac{\pi}{2}$. The LE over qubits $1$ and $2$ is obtained by performing local projection measurement in the basis of $\sigma^x_3$ on qubit $3$ in the GHZ state, leading to maximally entangled post-measurement states $\ket{\Phi^{\pm}}=\frac{1}{\sqrt{2}}(\ket{00}\pm\ket{11})$ on qubits $1$ and $2$, which subsequently leads to $E_{12}(\ket{\text{GHZ}}\bra{\text{GHZ}})=1$. On the other hand, $E_{12}(\ket{\text{gGHZ}}\bra{\text{gGHZ}})\leq E_{12}(\ket{\text{GHZ}}\bra{\text{GHZ}})$ for all values of $\alpha,\beta$. Our numerical analysis suggests that in the case of the three-qubit gGHZ states subjected to local noise, examples of both $E_{12}=E^\prime_{12}$ and $E_{12}>E_{12}^\prime$ exist, as discussed in the subsequent discussions. To compare the LE and the RLE for the class of generalized GHZ states, we specifically evaluate the absolute error $\varepsilon=E_{12}(\rho)-E_{12}^\prime(\rho)$. We find that in presence of noise on all the qubits or set of qubits,  $\varepsilon\sim 10^{-2}$. Fig.~\ref{fig:errors} depicts $\varepsilon$ as bit-flip and amplitude-damping noise is acting on the  qubits in the three-qubit gGHZ states. Note that for the PF noise, RLE can faithfully mimic LE with sufficiently low error ($\sim 10^{-3}$), which is not the case for the other types of noise considered in this paper. For the RLE under  PF noise, $E^\prime_{12}=E^\prime_{0,12}$, which implies that the optimal Pauli measurement on qubit $3$ is in  the basis of $\sigma^x$. 

Note that from the symmetry of the gGHZ state, 					
\begin{eqnarray}
\mathcal{E}_{12}(\rho_{13})&=&\mathcal{E}_{12}(\rho_{23}); \mathcal{E}_{12}(\rho_1)=\mathcal{E}_{12}(\rho_2),
\label{eq:symmetry_gghz}
\end{eqnarray} 
where $\mathcal{E}=E,E^\prime$, and for the RLE,  local projection measurement in the Pauli basis is always performed on qubit $3$. This modifies Eqs.~(\ref{eq:3hier_1})-(\ref{eq:3hier_4}) as 
\begin{eqnarray}
\label{eq:3hier_1g}
&&\max\left\{\mathcal{E}_{12}\left(\rho_{123}\right),\mathcal{E}_{12}\left(\rho_{12}\right)\right\}\nonumber \\ 
& \leq & \min\left\{\mathcal{E}_{12}\left(\rho_{13}\right),\mathcal{E}_{12}\left(\rho_{1}\right)\right\},\\
\label{eq:3hier_2g}
&&\max\left\{\mathcal{E}_{12}\left(\rho_{13}\right),\mathcal{E}_{12}\left(\rho_{1}\right)\right\}\leq  \mathcal{E}_{12}\left(\rho_3\right), \\
\label{eq:3hier_3g}
&&\mathcal{E}_{12}\left(\rho_{123}\right) \leq  \mathcal{E}_{12}\left(\rho_{12}\right),\\ 
\label{eq:3hier_4g}
&&\mathcal{E}_{12}\left(\rho_{13}\right) \leq \mathcal{E}_{12}\left(\rho_{1}\right).
\end{eqnarray} 
\noindent In the following \textbf{Propositions IV-V}, hierarchies among the values of RLE for  different values of $m$ and for different situations described in Sec.~\ref{sec:hierarchies} are discussed when single-qubit BF and PF noise are applied to the three-qubit gGHZ states. 

%The proofs of the \textbf{Propositions IV-VI} are given in Appendix~\ref{subsec:prop4}-\ref{subsec:prop6}. 

%$E^\prime_{12}(\rho_{12})$ and $E^\prime_{12}(\rho_{13})$ corresponding to $m=2$, and $E^\prime_{12}(\rho_{1})$ and $E^\prime_{12}(\rho_{3})$ corresponding to $m=1$ for different types of noise are discussed.  

\noindent\textbf{Proposition IV. Phase-flip channel.} \emph{The values of $E^\prime_{12}$ corresponding to the different values of the cardinality $m=1,2,3$, of the sets of  noisy qubits,  calculated by using negativity as the seed measure, over the qubits $1$ and $2$ of a three-qubit generalized GHZ state subjected to local uncorrelated phase-flip noise of strength $p$, $0\leq p\leq 1$, satisfy} 
\begin{eqnarray}
\label{eq:pf_hier_gghz}
E_{12}^\prime(\rho_{123})\leq E_{12}^\prime\left(\rho_{12}\right)&=&E_{12}^\prime\left(\rho_{13}\right)\nonumber \\
&\leq & E_{12}^\prime\left(\rho_{1}\right)=E_{12}^\prime\left(\rho_{3}\right).
\end{eqnarray}

\begin{figure*}
\includegraphics[width=\textwidth]{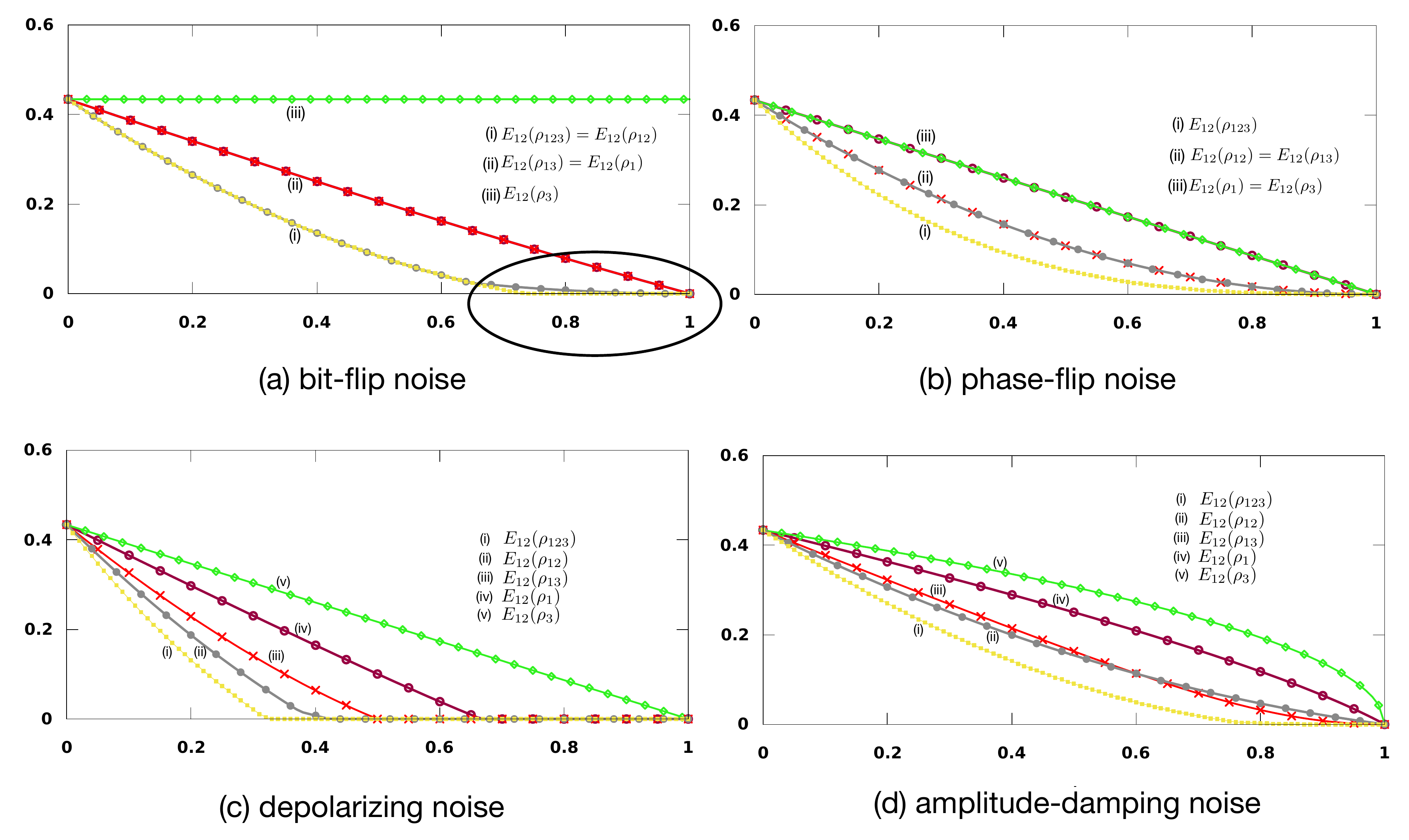}
\caption{Variations of the localizable entanglement (Y axis) as a function of $p$ (X axis) in the cases of bit-flip(a), phase-flip (b), depolarizing (c), and amplitude damping (d) noise acting locally on different sets of qubits in a three-qubit gGHZ state with $\alpha=\pi/3$, $\beta=0$. All quantities plotted are dimensionless.}
\label{fig:dynamics}
\end{figure*}

The proof of the proposition can be found in Appendix~\ref{app:proofs_IV}. When the bit-flip channel acts on the qubits, similar inequalities like \ref{eq:pf_hier_gghz} can be obtained by calculating localizable negativity in different scenarios. In particular, we have the following Proposition, the proof of which is given in Appendix~\ref{app:proofs_V}.

\noindent\textbf{Proposition V. Bit-flip channel.} \emph{When local bit-flip noise of strength  $p$, $0\leq p\leq 1$, acts on all or some of the qubits in a three-qubit generalized GHZ state, the RLE obey the following ranking.}
%\emph{The values of $E^\prime_{12}$ corresponding to the different values of the cardinality $m=1,2,3$, of the sets of  noisy qubits,  calculated by using negativity as the seed measure, over the qubits $1$ and $2$ of a three-qubit generalized GHZ state subjected to local uncorrelated bit-flip noise of strength $p$, $0\leq p\leq 1$, satisfy} 
\begin{eqnarray}
\label{eq:bf_hier}
E_{12}^\prime(\rho_{123})=E_{12}^\prime(\rho_{12})&\leq & E_{12}^\prime(\rho_{13})\nonumber\\ &=& E_{12}^\prime(\rho_1)\leq E_{12}^\prime(\rho_{3}). 
\end{eqnarray} 
%
%\noindent\textbf{Proof.} The case of the BF noise belongs to CASE 1 in the proof of \textbf{Proposition I}. Hence, by using \textbf{Corollary I.1},
%\begin{eqnarray}
%E_{12}^\prime(\rho_{123})&=&E^\prime_{12}(\rho_{12}),\nonumber \\
%E_{12}^\prime(\rho_{13})&=&E_{12}^\prime(\rho_1),\nonumber \\
%E_{12}^\prime(\rho_{23})&=&E_{12}^\prime(\rho_2).
%\end{eqnarray}
%Straightforward algebra leads to the expressions of $E_{12}^\prime(\rho_{12})$, $E_{12}^\prime(\rho_{1})$, and $E_{12}^\prime(\rho_{3})$ as 
%\begin{eqnarray}
%\label{eq:bfneg_1}
%E_{12}^\prime(\rho_{12})&=&\frac{1}{8}\left[\sin\alpha\sqrt{f}-2p(2-p)\right],\\
%\label{eq:bfneg_2}
%E_{12}^\prime(\rho_{1})&=&\frac{1}{4}\left[\sqrt{p^2+4(1-p)\sin^2\alpha}-p\right],\\
%\label{eq:bfneg_3}
%E_{12}^\prime(\rho_{3})&=&\frac{1}{2}\sin\alpha,
%\end{eqnarray}
%where 
%\begin{eqnarray}
%f&=&\left[p^2+(2-p)^2\right]^2-4p^2(2-p)^2\sin^2\beta. 
%\label{eq:bfneg_fun}
%\end{eqnarray}
%Note that at $p=0$, $E_{12}^\prime(\rho_{12})=E_{12}^\prime(\rho_{1})=E_{12}^\prime(\rho_{3})$. For increasing $p$ in the  range $0<p\leq 1$, $E_{12}^\prime(\rho_3)$ remains independent of $p$, while $E_{12}^\prime(\rho_{12})$ and $E_{12}^\prime(\rho_1)$ decreases monotonically with $p$. Also, for $p>0$, $E_{12}^\prime(\rho_{1})=0$ iff $p=1$ $\forall \alpha\neq 0$. This implies that $E_{12}^\prime(\rho_{1})\leq E_{12}^\prime(\rho_{3})$  for the full range of $p$, the equality being only at $p=0$. 

\begin{figure*}
\includegraphics[width=\textwidth]{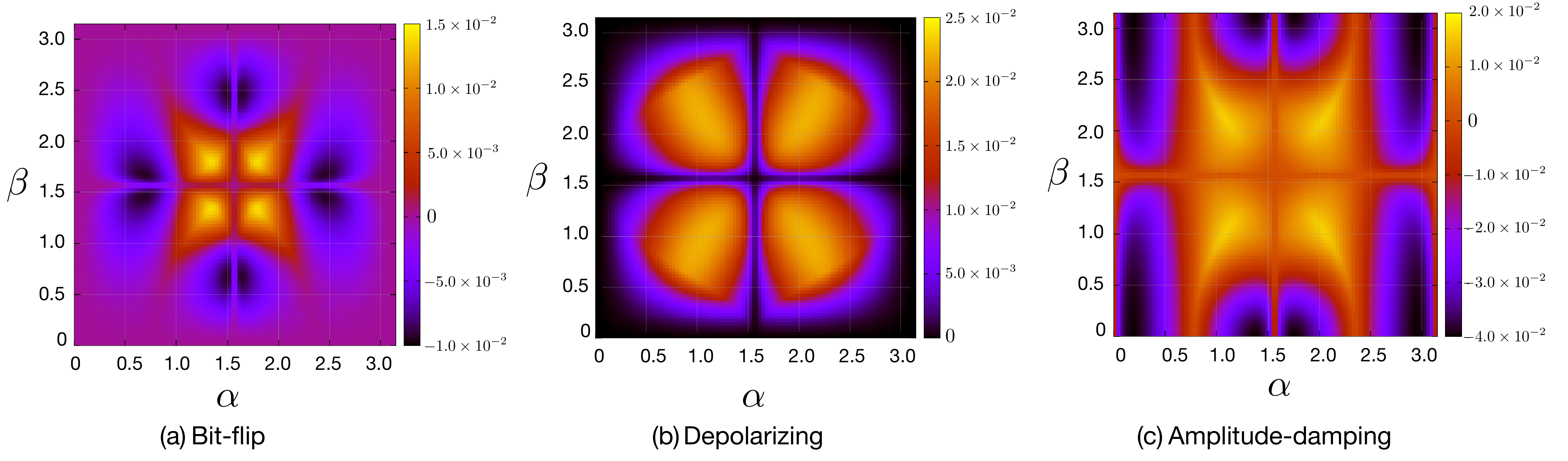}
\caption{(Colour online.) \textbf{Violation of hierarchy for gW states.} Variations of $\Delta_\text{B}$ (Eqs.~(\ref{eq:del_b})) for gW states as functions of $\alpha$ and $\beta$, with $\gamma_{1,2}=0$. Bit-flip  (a), depolarizing (b), and amplitude-damping (c) noise with $p=0.1$ acts on all the qubits. The parameters $\alpha$ and $\beta$ are in radians, while $\Delta_\text{B}$ is dimensionless.}
\label{fig:gw_violation}
\end{figure*}  
%
%On  the other hand, the function $E_{12}^\prime(\rho_{12})$ can be identified as $|\lambda|$, where $\lambda=\left[2p(2-p)-\sin\alpha\sqrt{f}\right]/8$ is the negative eigenvalue of the matrix obtained by performing partial transposition with respect to qubit $1$ on the post-measured state $\tilde{\rho}^{(x,k)}_{12}$ over qubits $1$ and $2$. Note that $\lambda$ is negative since  $\sin\alpha\sqrt{f}> 2p(2-p)$, which implies that $E_{12}^\prime(\rho_{12})=|\lambda|$ when 
%\begin{eqnarray}
%\frac{\left[p^2+(2-p)^2\right]^2}{4p^2(2-p)^2}>\frac{1+\sin^2\alpha\sin^2\beta}{\sin^2\alpha},
%\label{eq:neg_condition}
%\end{eqnarray}
%and $E_{12}^\prime(\rho_{12})=0$ otherwise. The condition (\ref{eq:neg_condition}) defines a critical value $p=p_c$ given by the solution of the equation obtained by replacing the inequality in  (\ref{eq:neg_condition}) by an equality, such that for $p< p_c$, $E_{12}^\prime(\rho_{12})=|\lambda|$, while for $p\geq p_c$, $E_{12}^\prime(\rho_{12})=0$. Therefore, for $p_c<p\leq 1$, $E^\prime_{12}(\rho_{12})\leq E^\prime_{12}(\rho_1)$, where use use the fact that $E_{12}^\prime(\rho_{1})=0$ iff $p=1$ $\forall \alpha\neq 0$.
%
%\noindent\textcolor{red}{Rest of the proof for the range $p<p_c$ is to be done!}\hfill $\blacksquare$

\noindent\textbf{Robustness of RLE.}  At this point, a word on the robustness of RLE of the gGHZ states under local uncorrelated Pauli noise is in order.  The robustness of the RLE can be quantified by the value of $p=p_c$ at which $E^\prime_{12}(\rho)$ vanishes. For the Markovian nature of the single-qubit uncorrelated Pauli noise, the value of $E^\prime_{12}(\rho)$ remains $0$ for $p\geq p_c$. The higher is the value of $p_c$, the more robust is the RLE for a specific set of noisy qubits. In the case of  the PF noise, the values of $E^\prime_{12}$ for all possible different sets $L$ of noisy qubits vanishes only at $p=1$. However, in  the case of the BF channel, $E^\prime_{12}(\rho_3)$ never goes to zero, while $E^\prime_{12}(\rho_{12})$ goes to zero at a  specific value $p=p_c\leq 1$, which is computed as the solution of the equation obtained from (\ref{eq:neg_condition}) by converting the inequality to an equality. The value of $p_c$ depends completely on the initial gGHZ state, and a value of $p_c<1$ implies a less robust behaviour of  $E^\prime_{12}(\rho_{12})$ compared to that of the other sets $L$ of qubits under the BF noise.

\noindent\textbf{Dynamics of LE.} It is now logical to ask whether the LE of the gGHZ states subjected to local noise on different sets of qubits obey the same hierarchies as the RLE. We anticipate  from Fig.~\ref{fig:dynamics} that  the answer can be negative. To support this view, in Fig.~\ref{fig:dynamics}, we consider examples of the variations of the LE as a function of the noise strength $p$, when local noise of BF, PF, DP, and AD types are applied to a set of chosen qubits in $\ket{\text{gGHZ}}$ with $\alpha=\frac{\pi}{3}$, $\beta=0$. It is clear from the variation of $E_{12}(\rho_{123})$ and $E_{12}(\rho_{12})$ with $p$ for the bit-flip noise that when the noise strength is high $(p\geq 0.7)$, $E_{12}(\rho_{123})< E_{12}(\rho_{12})$, although the maximum difference being very small (of the order of $10^{-2}$). This modifies the hierarchy obeyed by the LE, compared to the same for the RLE (Eq.~(\ref{eq:bf_hier})), as 
\begin{eqnarray}
\label{eq:bf_hier_le}
E_{12}(\rho_{123})<E_{12}(\rho_{12})&\leq & E_{12}(\rho_{13})\nonumber\\ &=& E_{12}(\rho_1)\leq E_{12}(\rho_{3}),
\end{eqnarray}
for high value of $p$. On the other hand, for the PF channel, the hierarchy for the RLE mimics the same for the LE with negligible error for the given example. 

Note, however, that the modified hierarchy in Eq.~(\ref{eq:bf_hier_le}) is still in accordance with the proposed hierarchies for three-qubit systems, as given in Eqs.~(\ref{eq:3hier_1})-(\ref{eq:3hier_4}). Our numerical findings suggest that the proposed hierarchies remain valid for gGHZ states.

%The first step towards investigating this is to check whether RLE faithfully represents LE for all values of $\alpha$, $\beta$, and $p$, for all types of noise considered in this paper. In Fig.~\ref{fig:errors}, we plot $\left(E_{12}-E_{12}^\prime\right)$ as functions of $p$, $\alpha$, and $\theta$, when bit-flip and depolarizing noise are applied to all the qubits, and amplitude-damping noise is applied to all qubits and qubits $1$ and $2$ only. Within our numerical precision, we set $E_{12}=E_{12}^\prime$ if $\left(E_{12}-E_{12}^\prime\right)\sim 10^{-3}$.  The values of $\left(E_{12}-E_{12}^\prime\right)$ in the color map clearly shows that there exists considerable fraction of three-qubit quantum states originated by applying local noise of the BF, DP, or AD type to a three-qubit gGHZ state for which  $\left(E_{12}-E_{12}^\prime\right)\sim 10^{-2}$, or higher. For the rest of the states, RLE can faithfully mimick LE with low-enough error, while for these states, one needs to check the hierarchies for the LE explicitly.    

\begin{table*}[t]
Random three-qubit states under phase-flip noise \\
\begin{tabular}{|c|c|c|c|}
\hline
& $p=0.1$ & $p=0.2$ & $p=0.3$  \\
\hline
          \begin{tabular}{c}
           State Type \\
           \hline
          GHZ Class \\
          \hline
          W Class \\
%          \hline
%          gW States \\
          \end{tabular}
          &
          \begin{tabular}{c|c|c}
            Env & A & B \\ 
          \hline 
           $98.78$  &$82.29$  &  $100.00$ \\
          \hline
         $98.55$  &$13.57$  &$100.00$   \\
%         \hline
%         $100.00$  & $100.00$ &$100.00$   \\
         \end{tabular}         
         &         
          \begin{tabular}{c|c|c}
           Env & A & B \\ 
          \hline 
          $98.95$  &$81.53$  &  $100.00$ \\
          \hline
          $98.37$  &$13.43$  &$100.00$   \\
%         \hline
%       $100.00$  & $100.00$ &$100.00$   \\
         \end{tabular}             
         &         
          \begin{tabular}{c|c|c}
            Env & A & B \\ 
          \hline 
         $99.07$  &$81.15$  &  $100.00$ \\
          \hline
        $98.29$  &$14.03$  &$100.00$   \\
%         \hline
%        $100.00$  & $100.00$ &$100.00$   \\
         \end{tabular}      \\   
\hline                      
\end{tabular}\\
Random three-qubit states under bit-flip noise\\
\begin{tabular}{|c|c|c|c|}
\hline
& $p=0.1$ & $p=0.2$ & $p=0.3$  \\
\hline
          \begin{tabular}{c}
            State type\\
           \hline
          GHZ Class \\
          \hline
          W Class \\
%          \hline
%          gW States  \\
          \end{tabular}
          &
          \begin{tabular}{c|c|c}
            Env & A & B \\ 
          \hline 
           $98.84$  &$82.56$  &  $99.99$ \\
          \hline
         $92.75$  &$58.23$  &$100.00$   \\
%         \hline
%         $90.03$  & $71.54$ &$100.00$   \\
         \end{tabular}         
         &         
          \begin{tabular}{c|c|c}
           Env & A & B \\ 
          \hline 
          $99.01$  &$81.87$  &  $99.99$ \\
          \hline
          $94.37$  &$58.94$  &$100.00$   \\
%         \hline
%       $91.57$  & $72.07$ &$100.00$   \\
         \end{tabular}             
         &         
          \begin{tabular}{c|c|c}
            Env & A & B \\ 
          \hline 
         $99.10$  &$81.42$  &  $100.00$ \\
          \hline
        $95.90$  &$59.46$  &$100.00$   \\
%         \hline
%        $93.47$  & $72.80$ &$100.00$   \\
         \end{tabular}       \\  
\hline                                             
\end{tabular}
\caption{\textbf{Percentage of three-qubit states under phase- and bit-flip noise, for which the proposed hierarchies for three-qubit systems are valid.} For each type of initial states, the sample size considered is $N_S=5\times 10^4$.}
\label{tab:3qubit_per}
\end{table*}

%
%Towards this purpose, we generate gGHZ states of the form given in Eq.~(\ref{eq:gGHZ}) by choosing the values of $\alpha$ and $\beta$ randomly from an uniform distribution in their allowed ranges, and apply local BF, PF, DP, and AD noise on different qubits. From the noisy states, LE is computed by performing an optimization over the real parameters $\theta$ and $\phi$. For demonstration, in 

\begin{table*}[t]
%\begin{tabular}{c}
Random 4 qubit states under phase-flip noise\\
\begin{tabular}{|c|c|c|}
\hline
 $p=0.1$ & $p=0.2$ & $p=0.3$  \\
\hline
          \begin{tabular}{c|c|c|c|c}
            $H_4$ & $H_5$ & $H_1$ & $H_2$ & $H_3$ \\ 
          
         \hline
          $63.72$  &  $31.82$  &$100.00$ &$90.41$ & $99.95$  \\
         
         \end{tabular}         
         &         
         \begin{tabular}{c|c|c|c|c}
            $H_4$ & $H_5$ & $H_1$ & $H_2$ & $H_3$ \\ 
          
         \hline
          $82.34$  &  $13.05$  & $100.00$ &$85.26$ & $99.98$  \\
          
         \end{tabular}             
         &         
          \begin{tabular}{c|c|c|c|c}
            $H_4$ & $H_5$ & $H_1$ & $H_2$ & $H_3$ \\ 
         
         \hline
          $92.602$  &  $4.19$  & $100.00$ &$80.83$ & $99.97$  \\
         
         \end{tabular}           \\  
       \hline  
         \end{tabular}\\
%         \end{tabular}
Random four-qubit states under bit-flip noise\\
\begin{tabular}{|c|c|c|}
\hline
 $p=0.1$ & $p=0.2$ & $p=0.3$  \\
\hline

          \begin{tabular}{c|c|c|c|c}
            $H_4$ & $H_5$ & $H_1$ & $H_2$ & $H_3$ \\ 
          \hline
          $63.85$ & $31.65$ & $100.00$ &$89.46$ & $99.59$ \\
         \end{tabular}         
         &         
         \begin{tabular}{c|c|c|c|c}
            $H_4$ & $H_5$ & $H_1$ & $H_2$ & $H_3$ \\ 
          
          \hline
          $82.87$ & $12.79$  & $100.00$ &$84.72$ & $99.82$
         \end{tabular}             
         &         
          \begin{tabular}{c|c|c|c|c}
            $H_4$ & $H_5$ & $H_1$ & $H_2$ & $H_3$ \\ 
          
          \hline
          $92.69$ & $4.34$ & $100.00$ &$80.66$ & $99.94$
         \end{tabular}           \\  
       \hline  
         \end{tabular}\\
\caption{\textbf{Percentage of four qubit states subjected to phase-flip and bit-flip noise,  for which the proposed hierarchies for the four-qubit systems are valid.} The sample size considered for each of the cases $N_S=5\times 10^4$.}
\label{tab:4qubit_per_pfbf}
\end{table*}

\noindent\textbf{gW states.} Let us now move to another class of three-qubit  states, namely, the generalized W states, whose parametric form is given by 
\begin{eqnarray}
\ket{\text{gW}}&=&\cos\alpha\ket{001}+\text{e}^{\text{i}\gamma_1}\sin\alpha\cos\beta\ket{010}\nonumber\\
&&+\text{e}^{\text{i}\gamma_2}\sin\alpha\sin\beta\ket{001}, 
\end{eqnarray}
where $0\leq\alpha,\beta\leq \pi$, and $0\leq\gamma_1,\gamma_2\leq 2\pi$. Due to the increased number of real parameters required for specifying the gW states, it is not possible to obtain analytical closed forms even for the RLE. We perform numerical analysis and find that similar to the gGHZ states, there exists gW states for which $E_{12}-E_{12}^\prime=\delta\sim 10^{-2}$ when even a low noise is applied to the qubits. This suggests that the hierarchies of LE and RLE have to be checked separately in the case of the gW states under noise.

Let us first check whether inequalities.~(\ref{eq:3hier_1})-(\ref{eq:3hier_4}) remain valid for LE in the case of gW states. We observe that the hierarchies labelled as ``Env" are valid for LE in all  states generated after interaction of  local BF and PF noise with the gW states. However, there can be violation of \textit{microscopic} hierarchies, labelled as ``A" and ``B". To consider the degree of violation of the hierarchy ``B", we consider a quantity  
%In Fig.~\ref{fig:gw_violation}, we plot $\Delta_\text{B}$ as functions of $\alpha$ and $\beta$, keeping $\gamma_{1,2}=0$, for the BF noise of strength $p=0.1$ for each qubit, where 
\begin{eqnarray}
%\label{eq:del_env}
%\Delta_{\text{Env}} &=& \min\left\{E_{12}(\rho_{13}),E_{12}(\rho_{23}),E_{12}(\rho_{1}),E_{12}(\rho_{2})\right\}\nonumber\\
%&&-\max\left\{E_{12}(\rho_{123}),E_{12}(\rho_{12})\right\},\\
%\label{eq:del_a}
%\Delta_\text{A} &=& E_{12}(\rho_{12})-E_{12}(\rho_{123}),\\
\label{eq:del_b}
\Delta_\text{B} &=& \min\left\{E_{12}(\rho_1),E_{12}(\rho_2)\right\}\nonumber\\
&&-\max\left\{E_{12}(\rho_{13}),E_{12}(\rho_{23})\right\}.
\end{eqnarray}
In Fig.~\ref{fig:gw_violation}, $\Delta_B$  is plotted with $\alpha$ and $\beta$, keeping $\gamma_{1,2}=0$, for the bit-flip noise strength $p=0.1$ for each qubit, and we notice that $\Delta_B$ is positive as well as negative, confirming the violation.  
%is the degrees of violations of the hierarchy ``B". 
%The subscripts of $\Delta$ refers to the different types of hierarchies introduced in the discussion succeeding Eqs.~(\ref{eq:3hier_1})-(\ref{eq:3hier_4}).  
%A negative value of $\Delta_\text{B}$ imply a violation of the corresponding hierarchy. From Fig.~\ref{fig:gw_violation}, it is clear that there exist substantial number of gW states for which the hierarchy labeled as ``B" is violated even when a low noise of  bit-flip type is applied on the system. Our 
Numerical analysis suggests that this feature remain qualitatively unchanged when the strengths of the noise is also increased. Note, however, that in the case of RLE, all the proposed hierarchies remain valid in the case of the gW states subjected to BF and PF noise of all possible noise-strength. 

\noindent\textbf{Random 3- and 4-qubit states.} 
%Note that the three-qubit states of the forms $\ket{\text{gGHZ}}$ and $\ket{\text{gW}}$ forms sets of measure zero in the set of all possible three-qubit pure states.  Since there exists three-qubit gGHZ and gW states for which the proposed hierarchies (Eqs.~(\ref{eq:3hier_1})-(\ref{eq:3hier_4})) are violated for LE (in noisy and/or noiseless scenario), it is now logical to 
Let us now investigate what is the fraction of states for which these hierarchies are violated in the case of three-qubit generic random pure states sent through local noisy channels. Towards this aim, we Haar-uniformly generate generic $3$-qubit pure states of the form  
\begin{eqnarray}
\ket{\psi}=\sum_{i_1i_2i_3=0,1}a_{i_1i_2i_3}\ket{i_1i_2i_3} 
\label{eq:3_qubit_random_states}
\end{eqnarray}
with $\sum_{i_1i_2i_3=0,1}|a_{i_1i_2i_3}|^2=1$ and $a_{i_1i_2i_3}=\alpha_{i_1i_2i_3}+\text{i}\beta_{i_1i_2i_3}$, where $\alpha_{i_1i_2i_3}$ and $\beta_{i_1i_2i_3}$ are real numbers, by choosing the values of $\alpha_{i_1i_2i_3}$ and $\beta_{i_1i_2i_3}$ from a Gaussian distribution of mean zero and standard deviation unity~\cite{bengtsson2006}. Here, $\ket{i_k}\in\{\ket{0},\ket{1}\}$, $k=1,2,3$, form the computational basis of qubits $1$, $2$, and $3$. These states form the GHZ class of three-qubit states~\cite{dur2000}. On the other hand, there exists another class of three-qubit states~\cite{dur2000}, called the W class of states, which can not  be transferred  to a  state from the  GHZ class by stochastic local operations and classical  communication with a single copy   A generic state belonging to the  W class is represented as 
\begin{eqnarray}
\ket{\psi}=a_0\ket{001}+a_1\ket{010}+a_2\ket{100}+a_3\ket{000}, 
\label{eq:3_qubit_w_class}
\end{eqnarray}
with $\sum_{l=0}^3|a_l|^2=1$, and $\{a_l;l=0,1,2,3\}$ being complex numbers $a_l=\alpha_l+\text{i}\beta_l$, $l=0,1,2,3$, with real $\alpha_l$ and $\beta_l$.  Similar to the GHZ class states, random W class states can be generated Haar  uniformly by  generating values of $\alpha_l$ and $\beta_l$, $j=0,1,2,3,$ from a normal distribution of mean zero and standard deviation unity. We Haar uniformly generate three-qubit states belonging to these two classes, and subject them to single-qubit local PF and BF noise on different qubits. The percentages of such states for which the hierarchies presented in Eqs.~(\ref{eq:3hier_1})-(\ref{eq:3hier_4}) remains valid are tabulated in Table~\ref{tab:3qubit_per}. The prominent observations from the data are as follows. 
\begin{itemize}
\item For both GHZ and W class states, the percentage of states for which hierarchies ``Env" and ``A"  remains valid varies very slowly with increasing noise strength for both the PF and BF noise. The maximum variation between any two fractions of such states, corresponding to any two different values of noise strengths, is $\sim 1\%$.   

\item The hierarchy ``B" remains valid for almost all states belonging to the GHZ and the  W classes under both PF and BF noise.  

\item The number of W class states for which the hierarchy ``A" is valid is considerably low in the case of the PF noise. The number increases in the case of the BF noise, but remains $\sim$ half of the number of states in W class for which the hierarchies ``Env" and ``A" are valid.  

\item We also observe that for the cardinality based hierarchy, all of the GHZ and the W class states subjected to local noise of BF or PF type are in agreement.  
\end{itemize}

In order to check whether similar trend exists for the four-qubit systems as well, we Haar uniformly generate four-qubit states of the form 
\begin{eqnarray}
\ket{\psi}=\sum_{i_1i_2i_3i_4=0,1}a_{i_1i_2i_3i_4}\ket{i_1i_2i_3i_4},
\label{eq:4_qubit_random_states}
\end{eqnarray}
where the coefficients $a_{i_1i_2i_3i_4}$, $i_1,i_2,i_3,i_4=0,1$, and the bases $\ket{i_k}$, $k=1,2,3,4,$ have similar implications as in the case for three qubits, and the complex state parameters $\{a_j\}$ are sampled in a way similar to that in the case of the 3-qubit GHZ and W class states. We follow the same labelling scheme for the hierarchies in four-qubit systems  as in Figs.~\ref{fig:4qubit_h13} and \ref{fig:4qubit_h45}, where $H_{4}$ and $H_5$ represent the envelope hierarchies. In Tables~\ref{tab:4qubit_per_pfbf}, we have tabulated the percentages of Haar uniform random four-qubit states that obey the hierarchies labelled as $H_1,\cdots,H_5$. Note here that while checking the hierarchy $H_2$, in order to obtain the broad picture instead of getting data cluttered with microscopic details, we have combined the four microscopic hierarchies into the following two:
\small 
\begin{eqnarray}
&&\max\left\{\mathcal{E}_{12(ii)}^{(1)}(\rho_{134}),\mathcal{E}_{12(ii)}^{(1)}(\rho_{234})\right\}  \nonumber \\ 
&\leq & \min\left\{\mathcal{E}_{12(ii)}^{(1)}(\rho_{13}),\mathcal{E}_{12(ii)}^{(1)}(\rho_{14}),\mathcal{E}_{12(ii)}^{(2)}(\rho_{23}),\mathcal{E}_{12(ii)}^{(2)}(\rho_{24})\right\},\\
&&\max\left\{\mathcal{E}_{12(ii)}^{(1)}(\rho_{13}),\mathcal{E}_{12(ii)}^{(1)}(\rho_{14}),\mathcal{E}_{12(ii)}^{(2)}(\rho_{23}),\mathcal{E}_{12(ii)}^{(2)}(\rho_{24})\right\}  \nonumber \\
&\leq & \min\left\{\mathcal{E}_{12(ii)}^{(1)}(\rho_{1}),\mathcal{E}_{12(ii)}^{(2)}(\rho_{2})\right\}, 
\end{eqnarray}  \normalsize
where we have kept both unmeasured qubits at the same footing. This is a logical choice when the states are generated Haar-uniformly in the space of four-qubit states, where noise on qubit $1$ is equivalent to noise on qubit $2$ in terms of statistics of the state space.

We now summarize the observations.  
\begin{itemize}
\item  The percentage of states for which the hierarchy $H_4$ (equivalent to the hierarchy ``Env" in the three-qubit system) remains valid is considerably low in the low-noise scenario (for example, $p=0.1$) in the case of both PF and BF noise, and increases with the increase in the noise strength. 

\item Almost all random Haar-uniform four-qubit states obey the \emph{microscopic} hierarchies $H_1$ and $H_3$, while the percentages of states obeying hierarchy $H_2$ is lower. Moreover, unlike the other \emph{microscopic} hierarchies for the four qubit states as well as the three-qubit system, the percentage of states obeying hierarchy $H_2$ decreases at a considerable rate with increase in the noise strength for both PF and BF noise. 

\item From our numerical data, it appears that the percentages of four-qubit states obeying hierarchy $H_4$ is approximately complementary to the fraction of four-qubit states for which hierarchy $H_5$ is valid. As the noise strength in the case of the PF and the BF noise increases, the fraction of states for which hierarchy $H_4$ ($H_5$) is valid decreases (increases). Note here that the hierarchy $H_5$ takes into account only the cardinality of the set of noisy qubits, and not the intricacies of localizable entanglement.  

%\item\textcolor{red}{A comparison of $H_5$ and the same for three-qubit system should be there, no?}  
\end{itemize}

\begin{table*}[ht]
%\begin{tabular}{c}
3-qubit states: Depolarizing noise\\
\begin{tabular}{|c|c|c|c|}
\hline
& $p=0.1$ & $p=0.2$ & $p=0.3$  \\
\hline
          \begin{tabular}{c}
           State Type \\
           \hline
          GHZ Class \\
          \hline
          W Class \\
%          \hline
%          gW states \\
          \end{tabular}
          &
          \begin{tabular}{c|c|c}
            Env & A & B \\ 
          \hline 
           $99.93$  &$100.00$  &  $100.00$ \\
          \hline
         $93.79$  &$100.00$  &$100.00$   \\
%         \hline
%          $87.25$  & $100.00$ &$100.00$   \\
         \end{tabular}         
         &         
          \begin{tabular}{c|c|c}
           Env & A & B \\ 
          \hline 
          $99.65$  &$100.00$  &  $100.00$ \\
          \hline
          $91.80$  &$100.00$  &$100.00$   \\
%         \hline
%        $87.65$  & $100.00$ &$100.00$   \\
         \end{tabular}             
         &         
          \begin{tabular}{c|c|c}
            Env & A & B \\ 
          \hline 
         $99.22$  &$100.00$  &  $100.00$ \\
          \hline
        $91.59$  &$100.00$  &$100.00$   \\
%         \hline
%         $88.76$  & $100.00$ &$100.00$   \\
         \end{tabular}  \\       

\hline          
\end{tabular}\\

4-qubit states: Depolarizing noise\\
\begin{tabular}{|c|c|c|}
\hline
$p=0.1$ & $p=0.2$ & $p=0.3$  \\
\hline
          
          \begin{tabular}{c|c|c|c|c}
            $H_4$ & $H_5$ & $H_1$ & $H_2$ & $H_3$ \\ 
         
          \hline
         $33.04$ & $62.23$ &$100.00$  &$100.00$ & $100.00$  \\

         \end{tabular}         
         &         
         \begin{tabular}{c|c|c|c|c}
                        $H_4$ & $H_5$ & $H_1$ & $H_2$ & $H_3$ \\ 
          
          \hline
         $51.05$ & $33.14$ &$100.00$  &$100.00$ & $100.00$  \\
       
         \end{tabular}             
         &         
          \begin{tabular}{c|c|c|c|c}
                        $H_4$ & $H_5$ & $H_1$ & $H_2$ & $H_3$ \\ 
         
          \hline
         $68.68$ & $8.236$ &$100.00$  &$100.00$ & $100.00$  \\

         \end{tabular}           \\  
       \hline  
         \end{tabular}\\
%\end{tabular}
\caption{Percentage of the three- and four-qubit states which, when sent through depolarizing channels, satisfy Eqs.~(\ref{eq:3hier_1})-(\ref{eq:3hier_4}) (for three qubits) and the hierarchies shown in Figs.~\ref{fig:4qubit_h13}-\ref{fig:4qubit_h45}. For each type of initial states, the sample size considered is $N_S=5\times 10^4$.}
\label{tab:dep}
\end{table*}

\section{Ordering of states affected by depolarizing channel}
\label{sec:dp}

In this section, we consider a symmetric noise model, namely, the DP noise, as opposed to the asymmetric BF and PF noise. Similar to the previous section, we start with the effect of DP noise on the three-qubit gGHZ state, and present the following \textbf{Proposition VI} for the RLE.

\noindent\textbf{Proposition VI. Depolarizing channel.} \emph{When all or a set of qubits of the generalized GHZ state are passed through DP channel, according to the value of  RLE, the following classification of states is possible.}
%\emph{The values of $E^\prime_{12}$ corresponding to the different values of the cardinality $m=1,2,3$, of the sets of  noisy qubits,  calculated by using negativity as the seed measure, over the qubits $1$ and $2$ of a three-qubit generalized GHZ state subjected to local uncorrelated depolarizing noise of strength $p$, $0\leq p\leq 1$, satisfy} 
\begin{eqnarray}
\label{eq:dp_hier}
E_{12}^\prime(\rho_{123})\leq E_{12}^\prime(\rho_{12})&\leq & E_{12}^\prime(\rho_{13})\nonumber\\
&\leq & E_{12}^\prime(\rho_1)\leq E_{12}^\prime(\rho_{3}).  
\end{eqnarray}

See Appendix~\ref{app:proofs_VI} for the proof. Note that in the case of the DP noise, at $p=0$, 
\begin{eqnarray}
E^\prime_{12}(\rho_{123})=E^\prime_{12}(\rho_{12})&=& E^\prime_{12}(\rho_{13})\nonumber \\ 
&=& E^\prime_{12}(\rho_{1})=E^\prime_{12}(\rho_{3}),
\end{eqnarray} 
all of which have maximum value $\frac{1}{2}\sin\alpha$ at $p=0$. For $0 < p\leq 1$, $E^\prime_{12}(\rho_3)$ decreases linearly with $p$, vanishing only at $p=1$, thereby showing a higher robustness compared  to the same for $E^\prime_{12}(\rho_{123})$, $E^\prime_{12}(\rho_{12})$, $E^\prime_{12}(\rho_{13})$ and $E^\prime_{12}(\rho_1)$, which decrease monotonically with $p$, and may vanish at $p=p_c\leq 1$.  The values of $p_c$ corresponding to $E^\prime_{12}(\rho_{123})$ and $E^\prime_{12}(\rho_{12})$ are respectively given by the solutions of the equations   
\begin{eqnarray}
E^\prime_{12}(\rho_{123}) &=&  0, E^\prime_{12}(\rho_{12}) =  0,
\end{eqnarray} 
%\begin{eqnarray}
%p_{c} &=&  1 - \frac{1}{\sqrt{1+2\sin\alpha}},\\
%p_{c} &=&  1 - \frac{1}{\sqrt{1+2\sin\alpha}},
%\end{eqnarray} 
which depends on the chosen initial gGHZ state via the parameter $\alpha$. However, in contrast, for $E^\prime_{12}(\rho_{13})$ and $E^\prime_{12}(\rho_{1})$, $p_c=\frac{1}{2}$ and $\frac{2}{3}$ respectively, which are independent of the chosen initial state.

Let us now check the validity of the hierarchies in LE in the case of the DP noise. For the gGHZ states, the proposed hierarchy of three-qubit states remain valid for LE, as demonstrated via the dynamics of the different LEs in Fig.~\ref{fig:dynamics}. However, in contrast to the BF noise, for the gW states, no evidence is found for the violation of any of the three-qubit hierarchies, as given in Eqs.~(\ref{eq:3hier_1})-(\ref{eq:3hier_4}). This is demonstrated in Fig.~\ref{fig:gw_violation} by the absence of negative values in the variation of $\Delta_\text{B}$ (Eq.~(\ref{eq:del_b})) against $\alpha$ and $\beta$ for the depolarizing noise with a specific noise strength.   

The trends of the fraction of randomly generated three- and four-qubit states for which the rankings are valid exhibit several contrasting behaviour to the same for the PF and the BF noise (see Table~\ref{tab:dep}). For example, in the case of the DP noise, both \emph{microscopic} hierarchies ``A" and ``B" are satisfied by all the randomly sampled states from the three-qubit GHZ and W classes, which is unlike the trend in the case of the PF and the BF noise. On the other hand, the number of states from these classes, for which the hierarchy ``Env" is valid, has a high value, which increases slowly with $p$, similar to the PF and the BF noise. This behaviour remains unchanged when the number of qubits is increased from $3$ to $4$, in the sense that all the \emph{microscopic} hierarchies, $H_1,H_2$, and $H_3$, are valid for $100\%$ of the randomly sampled Haar-uniform four-qubit states. However, the \emph{envelop} hierarchy $H_4$ is valid for a less number of states, and the fraction increases at a considerable rate when the noise strength is increased from $p=0.1$ to $p=0.3$. 

Note here that the variation of the size of population of four-qubit states, for which the cardinality-based hierarchy $H_5$ remains valid, also exhibits different trends from that of the PF and the BF noise. The complementary nature of the values of the fraction of states obeying $H_4$ and $H_5$ breaks down as the strength of the DP noise increases, although the individual behaviour of the percentages of four-qubit states obeying $H_4$ and $H_5$ against the noise strength remains qualitatively the same as in the case of PF and BF noise.

\begin{table*}[ht]
%\begin{tabular}{c}
3-qubit states: Amplitude-damping noise\\
\begin{tabular}{|c|c|c|c|}
\hline
& $p=0.1$ & $p=0.2$ & $p=0.3$  \\
\hline
          \begin{tabular}{c}
           State Type \\
           \hline
          GHZ Class \\
          \hline
          W Class \\
%          \hline
%          gW states \\
          \end{tabular}
          &
          \begin{tabular}{c|c|c}
           Env & A & B \\ 
          \hline 
           $99.08$  &$78.54$  &  $100.00$ \\
          \hline
         $73.19$  &$51.54$  &$100.00$   \\
%         \hline
%          $66.59$  & $51.10$ &$100.00$   \\
         \end{tabular}         
         &         
          \begin{tabular}{c|c|c}
           Env & A & B \\ 
          \hline 
          $98.90$  &$80.2$  &  $100.00$ \\
          \hline
          $73.43$  &$52.74$  &$100.00$   \\
%         \hline
%        $66.81$  & $52.23$ &$100.00$   \\
         \end{tabular}             
         &         
          \begin{tabular}{c|c|c}
            Env & A & B \\ 
          \hline 
         $98.53$  &$81.92$  &  $100.00$ \\
          \hline
        $73.85$  &$54.03$  &$100.00$   \\
%         \hline
%         $67.21$  & $53.46$ &$100.00$   \\
         \end{tabular}         \\
\hline                      
\end{tabular}\\
4 qubit states: Amplitude-damping Noise\\
\begin{tabular}{|c|c|c|c|}
\hline
 $p=0.1$ & $p=0.2$ & $p=0.3$  \\
\hline
        
          \begin{tabular}{c|c|c|c|c}
            $H_4$ & $H_5$ & $H_1$ & $H_2$ & $H_3$ \\ 
          \hline 
           $38.76$ & $34.35$ &$100.00$  &  $91.93$ & $100.00$\\         
         \end{tabular}         
         &         
         \begin{tabular}{c|c|c|c|c}
            $H_4$ & $H_5$ & $H_1$ & $H_2$ & $H_3$ \\ 
          \hline 
           $47.86$ & $23.26$ &$100.00$  &  $90.83$ & $100.00$\\         
         \end{tabular}             
         &         
          \begin{tabular}{c|c|c|c|c}
            $H_4$ & $H_5$ & $H_1$ & $H_2$ & $H_3$ \\ 
          \hline 
           $53.41$ & $16.92$ &$100.00$  &  $90.20$ & $100.00$\\   
         \end{tabular}           \\  
       \hline  
\end{tabular}
\caption{Percentage of three- and four-qubit states satisfying proposed rankings under amplitude-damping noise.  For each type of initial states, the sample size considered is $N_S=5\times 10^4$.}
\label{tab:ad}
\end{table*}

\section{Rankings of states induced by Amplitude-damping noise}
\label{sec:ad}

We now consider a non-Pauli noise, namely, the single-qubit uncorrelated AD noise. The first step is to investigate its effect on the hierarchies of the RLE. Interestingly,  in contrast with the cases of the Pauli noises considered in this paper,  in the case of the AD channel, the hierarchies are altered when one crosses a specific value $p=p_{cr}$ on the $p$-axis. The value of $p_{cr}$ is fully dependent on the initial state parameters. This is described by the following \textbf{Proposition VII}.  

\noindent\textbf{Proposition VII. Amplitude-damping channel.} \emph{For the gGHZ  states, the  effect of local  AD channel on all or a set of qubits may lead to  a ranking of RLE given by}
%\emph{The values of $E^\prime_{12}$ corresponding to the different values of the cardinality $m=1,2,3$, of the sets of  noisy qubits,  calculated by using negativity as the seed measure, over the qubits $1$ and $2$ of a three-qubit generalized GHZ state of the form given in Eq.~(\ref{eq:gGHZ}) subjected to local uncorrelated amplitude-damping noise of strength $p$, satisfy, in the full range $0\leq  p\leq 1$,} 
\begin{eqnarray}
\label{eq:ad_hier1}
E_{12}^\prime(\rho_{123})\leq E_{12}(\rho_{12})&\leq & E_{12}(\rho_{13})\nonumber \\ 
&\leq & E_{12}(\rho_1)\leq E_{12}(\rho_{3})
\end{eqnarray} 
\emph{for $\frac{\pi}{2} \leq\alpha \leq\pi$.  When $0\leq \alpha \leq \frac{\pi}{2}$,} 
\begin{eqnarray} 
\label{eq:ad_hier21}
E_{12}^\prime(\rho_{123})\leq E_{12}(\rho_{12})&\leq & E_{12}(\rho_{13})\nonumber \\ 
&\leq & E_{12}(\rho_1)\leq E_{12}(\rho_{3})
\end{eqnarray} 
\emph{for $0\leq p \leq p_{cr}$, and }
\begin{eqnarray} 
\label{eq:ad_hier22}
E_{12}^\prime(\rho_{123})\leq E_{12}(\rho_{12})&>& E_{12}(\rho_{13})\nonumber \\ 
&\leq & E_{12}(\rho_1)\leq E_{12}(\rho_{3})
\end{eqnarray} 
\emph{for $p_{cr}< p \leq 1$. Here, $p_{cr}$ is given by} 
\begin{eqnarray}
p_{cr} = \min\left[1,f(\alpha)\right],
\end{eqnarray}
\emph{where}
\begin{eqnarray}
f(\alpha)=\frac{2\sin \alpha-\sqrt{4\sin \alpha[\sin \alpha +\cos \alpha -1]}}{2(1-\cos\alpha)}.
\end{eqnarray}

Direct computation of all the  expressions lead to the inequalities as above, as  shown in Appendix~\ref{app:proofs_VII}.

Our numerical analysis shows that the crossing point $p=p_{cr}$ exists even for the case of LE ( $E_{12}(\rho_{12})$ and $\rho_{12}(\rho_{13})$) of gGHZ states subjected to AD noise, which has been demonstrated in Fig.~\ref{fig:dynamics}. While all of the three-qubit hierarchies (Eqs.~(\ref{eq:3hier_1})-(\ref{eq:3hier_4})) remain valid for the gGHZ states subjected to the AD noise, there exists a considerable number of gW states which, when subjected to AD noise, violates the \emph{microscopic} hierarchy ``B". In Fig.~\ref{fig:gw_violation}, a negative value of $\Delta_\text{B}$ demonstrates a violation of the hierarchy ``B", which is found in a considerable number of gW states, similar to the ones found in the case of the PF and the BF noise. Also, in the case of the three-qubit GHZ and W class states, all Haar-uniformly sampled states follow hierarchies ``B" -- a feature which remains constant qualitatively as well as quantitatively when the noise strength is increased. However, the fractions of the states obeying ``Env" and ``A" are lower, and varies slowly with $p$. In contrast, in the case of the four-qubit system, all the \emph{microscopic} hierarchies except $H_2$ remains valid for all random four-qubit states, irrespective of the strength of the noise. Also, the complementary nature of the fractions of states obeying the \emph{envelope} hierarchies $H_4$ and $H_5$ is lost in the present case, as in the case of DP noise. However, the individual trends of the fractions of random four-qubit states  obeying $H_4$ and $H_5$ against the noise strength remains qualitatively similar to that found in the case of the PF, BF, and the DP noise, i.e., percentage of states satisfying $H_4$ increases with the increase of $p$, while the same for $H_5$ decreases.

\section{Conclusions}
\label{sec:conclude}
A knowledge of how entanglement in a noisy quantum state is affected due to a spatial distribution of noise acting on the quantum state is essential for complete characterization of the state.  In this paper, we studied different orderings of the values of localizable and restricted localizable entanglement, computed using negativity as entanglement measure, over a specific qubit-pair in a multiqubit system, when local noise acts on the whole, or a group of qubits in the system. We proved that the information on the noise applied to a qubit in a multi-qubit state remains even after the local projection measurement and subsequent tracing out of the qubit from the system required to compute  localizable entanglement (LE) as long as the measurement-basis and the basis of the Kraus operators representing the noise model do not commute. This result remains unchanged for single-qubit phase-flip,  bit-flip, and depolarizing noise. Depending on these results, and the  properties of entanglement in noisy environments, and on our analytical results regarding the effect of single-qubit noise on the restricted localizable entanglement, we proposed a set of hierarchies that the value of the localizable or restricted localizable entanglement should obey when local noise acts on a set of qubits in the system.  We tested the proposed rankings among states based on localizable and restricted localizable entanglement in Haar uniformly generated random  three- and four-qubit systems. In the former, a number of paradigmatic classes of states, such as the generalized GHZ and the generalized W states, and the GHZ and the  W classes of states, are sent through local noisy channels of the mentioned types, and the percentages of states with respect to the validity of the proposed hierarchies are reported. 
%It is found that all the proposed hierarchies remain valid for the phase-flip and the depolarizing channels, while for the bit-flip and the amplitude-damping channels, the microscopic hierarchies may be violated. 
On the other hand,  we found that  a hierarchy that emerges simply from the cardinality of the set of noisy qubits  is violated for both three- and four-qubit states with the increase of  the strength of the noise. 
%We extend this study to the system of four qubits also, and find that almost all results remain qualitatively unchanged, while the fraction of random four-qubit states under noise, for which the proposed hierarchies remain valid, change considerably. 
Our results, therefore, opens up a new pathway for classifying the random states according to LE in the states-space 
%of $n$-qubit systems 
under the action of local noisy channels, and are expected to be useful in quantum information processing tasks like
% in scenarios where localizable entanglement is used as resource in a multiqubit system under local noise, such as 
%the cluster states in 
measurement-based quantum computations and the  quantum error-correcting codes in which states with nonvanishing LE  can be used as resources.

\acknowledgements

We acknowledge computations performed at the cluster computing facility of Harish-Chandra Research Institute, Allahabad, India and the use of \href{https://github.com/titaschanda/QIClib}{QIClib} -- a modern C++11 library for general purpose quantum computing. 

\appendix

\section{Projection measurement on gGHZ state under Pauli noise}
\label{app:proj}

Here, we discuss the effect of the local projection measurement in the basis of $\sigma^x$ on the third qubit of the three-qubit gGHZ state $\rho=\ket{\text{gGHZ}}\bra{\text{gGHZ}}$ under local uncorrelated Pauli noise, where we have adopted the notation in Figs.~\ref{fig:4qubit_h13}-\ref{fig:4qubit_h45} for the quantum states $\rho_3^m$. The explicit forms of the noisy states corresponding to the BF $(x)$, PF $(z)$, and DP $(xyz)$ noise are given by
\begin{eqnarray}
\label{eq:rho2xz}
\rho_{l_1l_2}^{(\delta)}&=&\left(1-\frac{p}{2}\right)^2\rho\nonumber \\ &&+\frac{p}{2}\left(1-\frac{p}{2}\right)\left[\sigma^{\delta}_{l_1}\rho\sigma^{\delta}_{l_1}+\sigma^{\delta}_{l_2}\rho\sigma^{\delta}_{l_2}\right]\nonumber\\
&&+\left(\frac{p}{2}\right)^2\sigma^{\delta}_{l_1}\sigma^{\delta}_{l_2}\rho\sigma^{\delta}_{l_1}\sigma^{\delta}_{l_2},\;\text{for}\;\delta=x,z,
\end{eqnarray}  
\begin{eqnarray}
\label{eq:rho1xz}
\rho_l^{(\delta)}&=&\left(1-\frac{p}{2}\right)\rho+\frac{p}{2}\sigma^{\delta}_l\rho\sigma^{\delta}_l,\;\text{for}\;\delta=x,z,
\end{eqnarray}
\begin{eqnarray}
\label{eq:rho2xyz}
\rho_{l_1l_2}^{(xyz)}&=&\left(1-\frac{3p}{4}\right)^2\rho\nonumber \\ 
&&+\frac{p}{4}\left(1-\frac{3p}{4}\right)\sum_{\delta=x,y,z}\left(\sigma^{\delta}_{l_1}\rho\sigma^{\delta}_{l_1}
+\sigma^{\delta}_{l_2}\rho\sigma^{\delta}_{l_2}\right)\nonumber \\ 
&&+\left(\frac{p}{4}\right)^2\sum_{\delta,\delta^\prime=x,y,z}\sigma_{l_1}^\delta\sigma_{l_2}^{\delta^\prime}\rho\sigma_{l_1}^\delta\sigma_{l_2}^{\delta^\prime},
\end{eqnarray} 
and 
\begin{eqnarray}
\label{eq:rho1xyz}
\rho_l^{(xyz)}&=&\left(1-\frac{3p}{4}\right)\rho+\frac{p}{4}\sum_{\delta=x,y,z}\sigma^{\delta}_l\rho\sigma^{\delta}_l,
\end{eqnarray} 
where $l_1\neq l_2$ and $l_1,l_2,l\in\{1,2,3\}$. The projection operators on qubit $3$ in the basis corresponding to a specific Pauli operator $\sigma^\gamma$, $\gamma=x,y,z$, are given by 
\begin{eqnarray}
P_{k}^\gamma=\frac{1}{2}\left(I+(-1)^k\sigma^\gamma\right),
\label{eq:pauli_projection}
\end{eqnarray}
where $k=0,1$ corresponds to the pair of outcomes. In the calculation, we shall use the following identity:
\begin{eqnarray}
\sigma^{\delta}P_k^\gamma\sigma^{\delta}=P_{k^\prime}^\gamma,
\label{eq:gghz_identity}
\end{eqnarray}
where $k^\prime=k$ if $\gamma=\delta$, and $k^\prime=k+1$ modulo $2$ if $\gamma\neq\delta$. Application of the projection operator $P^\gamma_k$ for $\gamma=x,y,z$, on any one of the qubits in the gGHZ state yields 
\begin{eqnarray}
P^\gamma_k\rho P^\gamma_k=\tilde{\rho}_k^{\gamma}\otimes P^\gamma_k,
\label{eq:gghz_measurement}
\end{eqnarray}
where $\tilde{\rho}^\gamma_k$ is defined on qubits $1$ and $2$. For $\gamma=x$,  
\begin{eqnarray}
\label{eq:rho12x}
\tilde{\rho}_k^{x}&=&\cos^2\frac{\alpha}{2}\ket{00}\bra{00}+\sin^2\frac{\alpha}{2}\ket{11}\bra{11}\nonumber \\
&&+(-1)^k\frac{1}{2}\sin\alpha\left[\text{e}^{-\text{i}\beta}\ket{00}\bra{11}+\text{e}^{\text{i}\beta}\ket{11}\bra{00}\right]\nonumber \\
\end{eqnarray}
for $k=0,1$. Note here that the superscript ``$x$" in $\tilde{\rho}_k^{x}$ indicates a projection measurement on qubit $3$ in the basis of $\sigma^x$. Using Eqs.~(\ref{eq:gghz_identity})-(\ref{eq:rho12x}), the noisy post-measured states on qubits $1$ and $2$ for different types of noise can be determined in different scenarios in a case-by-case basis.

\subsection{Noise on two qubits: \texorpdfstring{$m=2$}{m=2}}
We first consider the cases where noise is applied on any two of the three-qubit system (Eqs.~(\ref{eq:rho2xz}) and (\ref{eq:rho2xyz})). The possible situations constitute of two cases depending on whether noise is applied to the qubit $3$, on which the local projection measurement in the basis of $\sigma^x$ is performed. The two different situations are as follows.  

\subsubsection{\texorpdfstring{$l_1,l_2\in\{1,2\}$}{l1,l2 in {1,2}}}

Here, the third qubit is free of noise, leading to $k^\prime=k$ trivially from Eq.~(\ref{eq:gghz_identity}). Without any loss in generality, we assume $l_1=1$, $l_2=2$. Application of $P^x_k$ on the state $\rho_{l_1l_2}^{(\delta)}$ ($\delta=x,z$) leads to the post-measured reduced state $\tilde{\rho}_{12}^{(\delta,k)}$ on qubits $1$ and $2$ after tracing out qubit $3$, where
\begin{eqnarray}
\label{eq:two_noise_case_1}
\tilde{\rho}_{12}^{(\delta,k)}&=&\left(1-\frac{p}{2}\right)^2\tilde{\rho}^x_k
+\frac{p}{2}\left(1-\frac{p}{2}\right)\left[\sigma^{\delta}_{1}\tilde{\rho}^x_k\sigma^{\delta}_{1}+\sigma^{\delta}_{2}\tilde{\rho}^x_k\sigma^{\delta}_{2}\right]\nonumber \\
&&+\left(\frac{p}{2}\right)^2\sigma^{\delta}_{1}\sigma^{\delta}_{2}\tilde{\rho}^x_k\sigma^{\delta}_{1}\sigma^{\delta}_{2},
\end{eqnarray} 
with $\delta=x,z$, and $k=0,1$, which occurs with equal probability $p_k=\frac{1}{2}$. Note that the superscript $x$ in $\tilde{\rho}^x_{k}$ indicates the operation of $P^x_k$ on the initial gGHZ state, while the index $\delta$  $(\delta=x,z)$ in the superscript of $\tilde{\rho}_{12}^{(\delta,k)}$ denotes the type of noise (BF or PF) on the qubits $1$ and $2$. 

In the case of the DP noise on qubits $1$ and $2$ also, $k=k^\prime$ in Eq.~(\ref{eq:gghz_identity}) due to application of $p^x_k$ on qubit $3$ in $\rho_{l_1l_2}^{(xyz)}$, which leads to 
\begin{eqnarray}
\rho_{12}^{(xyz,k)}&=&\left(1-\frac{3p}{4}\right)^2\tilde{\rho}^x_k\nonumber \\
&&+\frac{p}{4}\left(1-\frac{3p}{4}\right)\sum_{\delta=x,y,z}\left(\sigma^{\delta}_{1}\tilde{\rho}^x_k\sigma^{\delta}_{1}+\sigma^{\delta}_{2}\tilde{\rho}^x_k\sigma^{\delta}_{2}\right)\nonumber \\ 
&&+\left(\frac{p}{4}\right)^2\sum_{\delta,\delta^\prime=x,y,z}\sigma_{1}^\delta\sigma_{2}^{\delta^\prime}\tilde{\rho}_k^x\sigma_{1}^\delta\sigma_{2}^{\delta^\prime},
\end{eqnarray}
after tracing out qubit $3$, where $\tilde{\rho}^x_k$ is given in Eq.~(\ref{eq:rho12x}).

\subsubsection{\texorpdfstring{$l_1=3$}{l1=3} or \texorpdfstring{$l_2=3$}{l2=3}}

In this situation, Eq.~(\ref{eq:gghz_identity}) dictates the post-measured quantum state. We assume $l_1=1,l_2=3$. In the case of BF noise ($\delta=x$), $k=k^\prime$ according to Eq.~(\ref{eq:gghz_identity}) for qubit $3$, and the post-measured reduced two-qubit state over qubits $1$ and $2$ is given by 
\begin{eqnarray}
\label{eq:two_noise_case_2}
\tilde{\rho}_{12}^{(x,k)}&=&\left(1-\frac{p}{2}\right)\tilde{\rho}^x_k+\frac{p}{2}\sigma_1^x\tilde{\rho}^x_k\sigma_1^x,
\end{eqnarray}
corresponding to $k=0,1$, which stand for the outcomes of the measurement with equal probability $p_k=\frac{1}{2}$. However, in the case of PF noise $(\delta=z)$, $k$ may not equal to $k^\prime$ on qubit $3$  for all the terms in the expansion of $\rho_{13}^{(z)}$ (Eq.~(\ref{eq:rho2xz})), and the application of $P^x_k$, $k=0,1$, on qubit $3$ in $\rho_{13}^{(z)}$ leads to
\begin{eqnarray}
\label{eq:two_noise_case_3_0}
\tilde{\rho}_{12}^{(z,0)}&=&\left(1-\frac{p}{2}\right)\left[\left(1-\frac{p}{2}\right)\tilde{\rho}^x_0+\frac{p}{2}\sigma_1^z\tilde{\rho}^x_0\sigma_1^z\right]\nonumber \\
&&+\frac{p}{2}\left[\left(1-\frac{p}{2}\right)\tilde{\rho}^x_1+\frac{p}{2}\sigma_1^z\tilde{\rho}^x_1\sigma_1^z\right],\\
\label{eq:two_noise_case_3_1}
\tilde{\rho}_{12}^{(z,1)}&=&\left(1-\frac{p}{2}\right)\left[\left(1-\frac{p}{2}\right)\tilde{\rho}^x_1+\frac{p}{2}\sigma_1^z\tilde{\rho}^x_1\sigma_1^z\right]\nonumber \\
&&+\frac{p}{2}\left[\left(1-\frac{p}{2}\right)\tilde{\rho}^x_0+\frac{p}{2}\sigma_1^z\tilde{\rho}^x_0\sigma_1^z\right].
\end{eqnarray}

In the case of the DP noise, similar situations as in the case of the PF noise arise, and the post-measured states on qubits $1$ and $2$ corresponding to the projection measurement outcomes $k=0,1$ due to the application of $P^x_k$ on qubit $3$ are given by 
\begin{eqnarray}
\label{eq:two_noise_case_4_0}
\tilde{\rho}_{12}^{(xyz,0)}&=&\left(1-\frac{p}{2}\right)\left[\left(1-\frac{3p}{4}\right)\tilde{\rho}^x_0+\frac{p}{4}\sum_{\delta=x,y,z}\sigma_1^{\delta}\tilde{\rho}^x_0\sigma_1^{\delta}\right]\nonumber \\
&&+\frac{p}{2}\left[\left(1-\frac{3p}{4}\right)\tilde{\rho}^x_1+\frac{p}{4}\sum_{\delta=x,y,z}\sigma_1^{\delta}\tilde{\rho}^x_1\sigma_1^{\delta}\right],\\
\label{eq:two_noise_case_4_1}
\tilde{\rho}_{12}^{(xyz,1)}&=&\left(1-\frac{p}{2}\right)\left[\left(1-\frac{3p}{4}\right)\tilde{\rho}^x_1+\frac{p}{4}\sum_{\delta=x,y,z}\sigma_1^{\delta}\tilde{\rho}^x_1\sigma_1^{\delta}\right]\nonumber \\
&&+\frac{p}{2}\left[\left(1-\frac{3p}{4}\right)\tilde{\rho}^x_0+\frac{p}{4}\sum_{\delta=x,y,z}\sigma_1^{\delta}\tilde{\rho}^x_0\sigma_1^{\delta}\right].
\end{eqnarray}

\subsection{Noise on a single qubit: \texorpdfstring{$m=1$}{m=1}}
We now focus on the case Pauli noise applied to a single quit in the three-qubit system  (Eqs.~(\ref{eq:rho1xz}) and (\ref{eq:rho1xyz})). Similar to the case of $m=2$, here also exist two different situations, as follows. 

\subsubsection{\texorpdfstring{$l\neq3$}{l is not 3}}

In situations where qubit $3$ is free of noise, the post-measurement two-qubit reduced state on qubits $1$ and $2$ corresponding to the different outcomes $k$ of the projection measurement $P^x_k$ on qubit $3$ for different types of noise are as follows. 

\noindent\textbf{BF/PF noise:} 

\begin{eqnarray}
\label{eq:rho12xz}
\rho_{12}^{(\delta,k)}&=&\left(1-\frac{p}{2}\right)\tilde{\rho}_k^x+\frac{p}{2}\sigma^{\delta}_l\tilde{\rho}_k^x\sigma^{\delta}_l,\;\;\delta=x,z.
\end{eqnarray}

\noindent\textbf{DP noise:}

\begin{eqnarray}
\label{eq:rho12xyz}
\rho_{12}^{(xyz,k)}&=&\left(1-\frac{3p}{4}\right)\tilde{\rho}_k^x+\frac{p}{4}\sum_{\delta=x,y,z}\sigma^{\delta}_l\tilde{\rho}_k^x\sigma^{\delta}_l.
\end{eqnarray}

\subsubsection{\texorpdfstring{$l=3$}{l is 3}}

In situations where qubit $3$ is noisy, the post-measurement two-qubit reduced state on qubits $1$ and $2$ corresponding to the different outcomes $k$ of the projection measurement $P^x_k$ on qubit $3$ for different types of noise are as follows. 

\noindent\textbf{BF noise:} 

\begin{eqnarray}
\label{eq:rho12xz_2}
\rho_{12}^{(x,k)}&=&\tilde{\rho}_k^x.
\end{eqnarray}

\noindent\textbf{PF and DP noise:} 

\begin{eqnarray}
\label{eq:rho12xz_3_0}
\rho_{12}^{(xyz,0)}=\rho_{12}^{(z,0)}&=&\left(1-\frac{p}{2}\right)\tilde{\rho}_0^x+\left(\frac{p}{2}\right)\tilde{\rho}_1^x,\\
\rho_{12}^{(xyz,1)}=\rho_{12}^{(z,1)}&=&\left(1-\frac{p}{2}\right)\tilde{\rho}_1^x+\left(\frac{p}{2}\right)\tilde{\rho}_0^x.
\label{eq:rho12xz_3_1}
\end{eqnarray}

\section{Proofs of the Propositions}
\label{app:proofs}

\subsection{Proof of Proposition IV} 
\label{app:proofs_IV}
We first consider the case of $m=3$. A local projection measurement $P^x_{k}$ on qubit $3$ in the basis of $\sigma^x$ leads to the post-measurement states $\tilde{\rho}_{12}^{(z,k)}$, $k=0,1$ (see Appendix~\ref{app:proj}). The probabilities $p_k$ of getting the measurement outcomes $k=0,1$ on qubit $3$ are the same $(p_k=1/2, k=0,1)$. The matrices obtained by performing partial transposition with respect to qubit $1$ in the states $\tilde{\rho}_{12}^{(z,0)}$ and $\tilde{\rho}_{12}^{(z,1)}$ have identical eigenvalues, given by
\begin{eqnarray}
\lambda_1 &=&\frac{1}{2}\left(1+\cos\alpha\right),\lambda_2 =\frac{1}{2}\left(1-\cos\alpha\right),\nonumber \\
\lambda_3 &=&\frac{1}{2}\left(1-p\right)^3\sin\alpha, \lambda_4 =-\frac{1}{2}\left(1-p\right)^3\sin\alpha,
\end{eqnarray}  
with $p$ $(0\leq p\leq 1)$ being the single-qubit noise-strength. From the definition of the gGHZ state,  one can assume $\cos^2\frac{\alpha}{2}\geq\sin^2\frac{\alpha}{2}$ without any loss in generality, which imposes the restriction $0\leq\alpha\leq \frac{\pi}{2}$ on $\alpha$. In this region, $\lambda_4<0$ while  $\lambda_{1,2,3}\geq 0$. Therefore, the localizable negativity is given by 
\begin{eqnarray} 
E_{12}^\prime(\rho_{123})=\frac{1}{2}(1-p)^3\sin\alpha.
\end{eqnarray}  
Proceeding in similar fashion, it can be shown that in the case of $m=2$,  
\begin{eqnarray} 
E_{12}(\rho_{12})=E_{12}(\rho_{13})=\frac{1}{2}(1-p)^2\sin\alpha, 
\end{eqnarray}  
while in the case of $m=1$, 
\begin{eqnarray} 
E_{12}(\rho_{l})=\frac{1}{2}(1-p)\sin\alpha\;\forall l=1,2,3.
\end{eqnarray}   
Noticing the power of $(1-p)$ in the expressions for RLEs,  
\begin{eqnarray}
\label{eq:rle_pf_hier}
E_{12}^\prime(\rho_{123})\leq E_{12}^\prime\left(\rho_{12}\right)&=&E_{12}^\prime\left(\rho_{13}\right)\nonumber \\
&\leq & E_{12}^\prime\left(\rho_{1}\right)=E_{12}^\prime\left(\rho_{3}\right).
\end{eqnarray}
Hence the proof.  \hfill $\blacksquare$

\subsection{Proof of Proposition V}
\label{app:proofs_V}

The case of the BF noise belongs to CASE 1 in the proof of \textbf{Proposition I}. Hence, by using \textbf{Corollary I.1},
\begin{eqnarray}
E_{12}^\prime(\rho_{123})&=&E^\prime_{12}(\rho_{12}),\nonumber \\
E_{12}^\prime(\rho_{13})&=&E_{12}^\prime(\rho_1),\nonumber \\
E_{12}^\prime(\rho_{23})&=&E_{12}^\prime(\rho_2).
\end{eqnarray}
Straightforward algebra leads to the expressions of $E_{12}^\prime(\rho_{12})$, $E_{12}^\prime(\rho_{1})$, and $E_{12}^\prime(\rho_{3})$ as 
\begin{eqnarray}
\label{eq:bfneg_1}
E_{12}^\prime(\rho_{12})&=&\frac{1}{8}\left[\sin\alpha\sqrt{f}-2p(2-p)\right],\\
\label{eq:bfneg_2}
E_{12}^\prime(\rho_{1})&=&\frac{1}{4}\left[\sqrt{p^2+4(1-p)\sin^2\alpha}-p\right],\\
\label{eq:bfneg_3}
E_{12}^\prime(\rho_{3})&=&\frac{1}{2}\sin\alpha,
\end{eqnarray}
where 
\begin{eqnarray}
f&=&\left[p^2+(2-p)^2\right]^2-4p^2(2-p)^2\sin^2\beta. 
\label{eq:bfneg_fun}
\end{eqnarray}
Note that at $p=0$, $E_{12}^\prime(\rho_{12})=E_{12}^\prime(\rho_{1})=E_{12}^\prime(\rho_{3})$. For increasing $p$ in the  range $0<p\leq 1$, $E_{12}^\prime(\rho_3)$ remains independent of $p$, while $E_{12}^\prime(\rho_{12})$ and $E_{12}^\prime(\rho_1)$ decreases monotonically with $p$. Also, for $p>0$, $E_{12}^\prime(\rho_{1})=0$ iff $p=1$ $\forall \alpha\neq 0$. This implies that $E_{12}^\prime(\rho_{1})\leq E_{12}^\prime(\rho_{3})$  for the full range of $p$, the equality being only at $p=0$.

\begin{figure}
\includegraphics[scale=0.5]{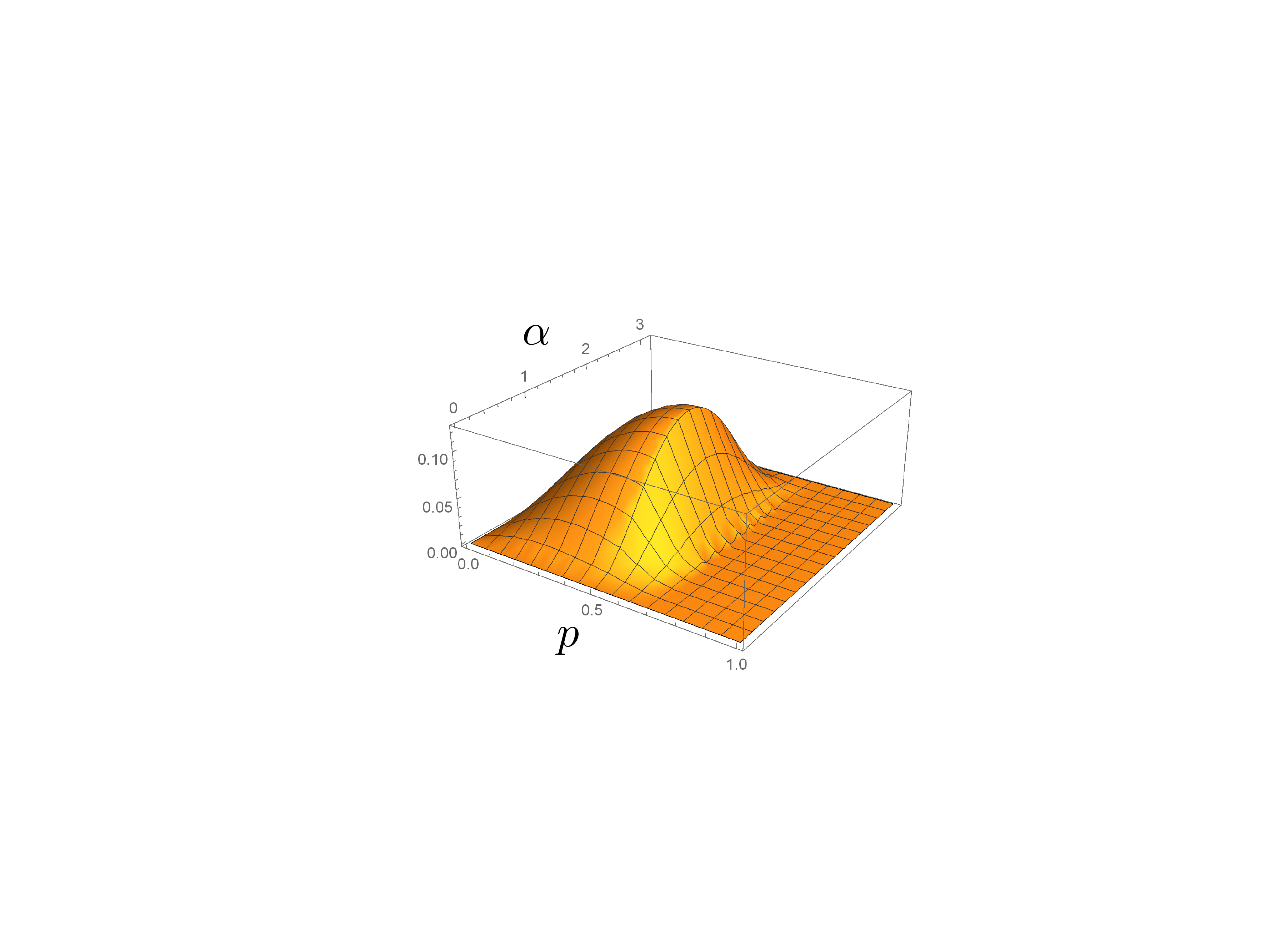}
\caption{(Colour online) Variation of $E_{12}^\prime(\rho_{1}) - E_{12}^\prime(\rho_{13})>0$ as a function of $\alpha$ and $p$,  where $\alpha$ is in radian, and $p$ is dimensionless, so is the y-axis.}
\label{fig:appendix}
\end{figure}

On  the other hand, the function $E_{12}^\prime(\rho_{12})$ can be identified as $|\lambda|$, where $\lambda=\left[2p(2-p)-\sin\alpha\sqrt{f}\right]/8$ is the negative eigenvalue of the matrix obtained by performing partial transposition with respect to qubit $1$ on the post-measured state $\tilde{\rho}^{(x,k)}_{12}$ over qubits $1$ and $2$. Note that $\lambda$ is negative since  $\sin\alpha\sqrt{f}> 2p(2-p)$, which implies that $E_{12}^\prime(\rho_{12})=|\lambda|$ when 
\begin{eqnarray}
\frac{\left[p^2+(2-p)^2\right]^2}{4p^2(2-p)^2}>\frac{1+\sin^2\alpha\sin^2\beta}{\sin^2\alpha},
\label{eq:neg_condition}
\end{eqnarray}
and $E_{12}^\prime(\rho_{12})=0$ otherwise. The condition (\ref{eq:neg_condition}) defines a critical value $p=p_c$ given by the solution of the equation obtained by replacing the inequality in  (\ref{eq:neg_condition}) by an equality, such that for $p< p_c$, $E_{12}^\prime(\rho_{12})=|\lambda|$, while for $p\geq p_c$, $E_{12}^\prime(\rho_{12})=0$. Therefore, for $p_c<p\leq 1$, $E^\prime_{12}(\rho_{12})\leq E^\prime_{12}(\rho_1)$, where use use the fact that $E_{12}^\prime(\rho_{1})=0$ iff $p=1$ $\forall \alpha\neq 0$. On the other hand, in the range $0<p<p_c$, we observe that prooving $E^\prime_{12}(\rho_{12})\leq E^\prime_{12}(\rho_1)$ is equivalent to proving $E^\prime_{12}(\rho_1)-\max\left[E^\prime_{12}(\rho_{12})\right]>0$. Noting that $E^\prime_{12}(\rho_{12})$ is maximum at $\beta=0,\pi,2\pi,\cdots,$ one can show that  $E^\prime_{12}(\rho_1)-\max\left[E^\prime_{12}(\rho_{12})\right]>0$ in the range $0<p<p_c$, thereby completing the proof. \hfill $\blacksquare$

\subsection{Proof of Proposition VI}
\label{app:proofs_VI}

Following the same prescription as in Appendix~\ref{app:proj} and the proofs of the \textbf{Propositions IV-V},   the expressions for $E^\prime_{12}(\rho_{123})$,  $E^\prime_{12}(\rho_{12})$, $E^\prime_{12}(\rho_{13})$, $E^\prime_{12}(\rho_{1})$ and $E^\prime_{12}(\rho_{3})$ are calculated as
\begin{eqnarray}
\label{eq:dpneg_1}
E^\prime_{12}(\rho_{123})&=&\frac{1}{4}\left[2(1-p)^3\sin\alpha - (2-p)p\right],\\
\label{eq:dpneg_2}
E^\prime_{12}(\rho_{12})&=&\frac{1}{4}\left[2(1-p)^2\sin\alpha - (2-p)p\right],\\
\label{eq:dpneg_3}
E^\prime_{12}(\rho_{13})&=&\frac{1}{8}\left[\sqrt{4p^2 +f_1}-2p\right],\\
\label{eq:dpneg_4}
E^\prime_{12}(\rho_{1})&=&\frac{1}{8}\left[\sqrt{4p^2 + f_2}-2p\right],\\
\label{eq:dpneg_5}
E^\prime_{12}(\rho_{3})&=&\frac{1}{2}\left[(1-p)\sin\alpha\right],
%\label{eq:dpneg_5}
%E_{12}(\rho_{123})&=&\frac{1}{4}\left[ 2\sin\alpha(1-p)^3 -(2-p)p \right] ,
\end{eqnarray}
where 
\begin{eqnarray}
f_1&=& 2\sin^2\alpha\left(8-32p+46p^2 +8p^4-32p^3\right),\\
f_2&=& \sin^2\alpha(p-2)(3p-2).
\label{eq:dpneg_fun}
\end{eqnarray}
It is clear from the expressions of $E^\prime_{12}(\rho_{123})$ and $E^\prime_{12}(\rho_{12})$ that $E^\prime_{12}(\rho_{123})\leq E^\prime_{12}(\rho_{12})$  for the full range of $p$ $\forall \alpha$. Also, straightforward algebra shows that $E^\prime_{12}(\rho_{13})- E^\prime_{12}(\rho_{12})$ can be simplified as 
%\begin{widetext}
\begin{eqnarray}
E^\prime_{12}(\rho_{13})- E^\prime_{12}(\rho_{12})=\nonumber\\ \frac{1}{8}\left[\sqrt{4p^2 +f_1} -2(1-p)\left(2(1-p)\sin\alpha - p\right)\right]
%4p^3(2-p)+16\sin\alpha p(1-p)^3 \nonumber \\
%&&+4\sin^2\alpha\Big[3 - 12p \nonumber \\
%&&++17p^2 -12p^3 +3p^4],
\end{eqnarray} 
%\end{widetext}
Simple algebra follows,
\begin{eqnarray}
(4p^2 +f_1) - 4(1-p)^2\left(2(1-p)\sin\alpha - p\right)^2 &=& \nonumber\\ 4p^3(2-p)+16\sin\alpha p(1-p)^3 \nonumber \\
+4\sin^2\alpha\Big[3 - 12p +17p^2 -12p^3 +3p^4\Big]
\end{eqnarray}
which is a positive quantity for the full range of  $p$, and for the allowed values of $\alpha$, i.e., $(0\leq\alpha\leq\frac{\pi}{2})$,  implying $E^\prime_{12}(\rho_{13})\geq E^\prime_{12}(\rho_{12})$. Similarly, for $E^\prime_{12}(\rho_{3})$ and $E^\prime_{12}(\rho_1)$,   
\begin{eqnarray}
E^\prime_{12}(\rho_{3})- E^\prime_{12}(\rho_{1})&=&\frac{p}{2}\sin\alpha\left[4(1-p)+p\sin\alpha\right],
\end{eqnarray}
which is $>0$ for all values of $p$ and allowed values of $\alpha$, thereby proving $E^\prime_{12}(\rho_{3})\geq E^\prime_{12}(\rho_{1})$.  On the other hand, $E_{12}^\prime(\rho_{1}) - E_{12}^\prime(\rho_{13})>0$ for all values of $\alpha$, as shown in Fig.~\ref{fig:appendix}. Hence the proof. \hfill $\blacksquare$

\subsection{Proof of Proposition VII}
\label{app:proofs_VII}

The expressions of $E_{12}(\rho_{12})$, $E_{12}(\rho_{13})$,$E_{12}(\rho_{1})$ and $E_{12}(\rho_{3})$ are 
\begin{eqnarray}
\label{eq:adneg_1}
E_{12}^\prime(\rho_{12})&=&\frac{1}{2}\left[(1-p)(\sin\alpha +p\cos\alpha -p)\right],\\
\label{eq:adneg_2}
E_{12}^\prime(\rho_{13})&=&\frac{1}{4}\left[\sqrt{f_1 + 4 p^2 \sin^4\frac{\alpha}{2}}-2p \sin^2\frac{\alpha}{2}\right],\\
\label{eq:adneg_3}
E_{12}^\prime(\rho_{1})&=&\frac{1}{8}\left[\sqrt{  f_2+4 p^2 \sin^4\frac{\alpha}{2}}-2p \sin^2\frac{\alpha}{2}\right],\\
\label{eq:adneg_4}
E_{12}^\prime(\rho_{3})&=&\frac{1}{2}\left[\sqrt{1-p}\sin \alpha\right],\\
\label{eq:adneg_5}
E_{12}^\prime(\rho_{123})&=&\frac{1}{2}\left[ \sqrt{(1-p)^3 \sin^2 \alpha} - p(1-p)(1-\cos \alpha) \right],\nonumber \\
\end{eqnarray}
where 
\begin{eqnarray}
f_1&=& 4(1-p)^2\sin^2 \alpha,\\
f_2&=& 4(1-p)\sin^2 \alpha .
\label{eq:adneg_fun}
\end{eqnarray}
Note that at $p=0$, $E_{12}^\prime(\rho_{12})=E_{12}^\prime(\rho_{13})=E_{12}^\prime(\rho_{1})=E_{12}^\prime(\rho_{3})$ and the maximum value of these quantities occur at $\alpha=\pi/2$. In the range $0< p\leq 1$, $E_{12}(\rho_3)$ , $E_{12}^\prime(\rho_{13})$ and $E_{12}^\prime(\rho_1)$ decrease monotonically with increasing $p$, and vanish only at $p=1$. In contrast,  $E_{12}^\prime(\rho_{12})$ may vanish at a critical value $p=p_c$, which depends on the state parameter $\alpha$, and is given by   
\begin{eqnarray}
p_c &=&\min\left[\cot\frac{\alpha}{2},1\right].
\end{eqnarray}
In the range $0\leq \alpha \leq \frac{\pi}{2}$, $\cot\frac{\alpha}{2}>1$,  implying $p_c=1$. On the other hand, in the range  $\frac{\pi}{2}\leq \alpha \leq \pi$, $\cot\frac{\alpha}{2}<1$,  leading to $p_c=\cot\frac{\alpha}{2}$. 

In the case of $E_{12}^\prime(\rho_{13})$  and $E_{12}^\prime(\rho_{1})$, note that both $f_1, f_2>0$  for $0\leq p\leq 1$,  and $f_2>f_1$ $\forall p$, thereby leading to $E_{12}^\prime(\rho_{1})\geq E_{12}^\prime(\rho_{13})$. The values of these quantities vanish only at  $p=1$.  

Next, we consider the difference between $E_{12}^\prime(\rho_{13})$ and $E_{12}^\prime(\rho_{12})$ as
\begin{eqnarray}
&&E_{12}^\prime(\rho_{13})- E_{12}^\prime(\rho_{12})\nonumber \\ 
&=&\frac{1}{4}\left(\sqrt{4(1-p)^2\sin^2\alpha +4p^2 \sin^4\frac{\alpha}{2} }  - g\right), 
\end{eqnarray}
where
\begin{eqnarray}
g = (4p^2 \sin^2\frac{\alpha}{2}+ 2\sin \alpha-2p \sin^2\frac{\alpha}{2}-2p\sin \alpha). 
\end{eqnarray}
Note that the solution of $p$ from the equation $E_{12}^\prime(\rho_{13})- E_{12}^\prime(\rho_{12})=0$ provides a crossing point of the curves representing  the variations of $E_{12}^\prime(\rho_{13})$ and $E_{12}^\prime(\rho_{12})$, which is given by 
\begin{eqnarray}
p_{cr} = \min\left[1,f(\alpha)\right],
\end{eqnarray}
where 
\begin{eqnarray}
f(\alpha)=\frac{2\sin \alpha-\sqrt{4\sin \alpha[\sin \alpha +\cos \alpha -1]}}{2(1-\cos \alpha)}. 
\end{eqnarray}
In the range $0\leq \alpha \leq \frac{\pi}{2}$, $0\leq p < p_{cr}$ ($p_{cr}< p < 1$), $E_{12}^\prime(\rho_{13})>E_{12}^\prime(\rho_{12})$ $\left(E_{12}^\prime(\rho_{13})<E_{12}^\prime(\rho_{13})\right)$. On the other hand, in the range  $\frac{\pi}{2} \leq\alpha \leq\pi$,  $E_{12}^\prime(\rho_{13})- E_{12}^\prime(\rho_{12})=0$ only at $p =1$, and  $E_{12}^\prime(\rho_{13})>E_{12}^\prime(\rho_{12})$ for the whole range of $p$.

Next, for  $E^\prime_{12}(\rho_{1})$ and $E^\prime_{12}(\rho_{3})$,  we get 
\begin{eqnarray}
&& E_{12}^\prime(\rho_{3})- E_{12}^\prime(\rho_{1})\nonumber\\
&=& \frac{1}{4}\left(h -\sqrt{ 4(1-p)\sin^2 \alpha +4p^2 \sin^4\frac{\alpha}{2} } \right),
\label{eq:positive_condition}
\end{eqnarray}
where 
\begin{eqnarray}
h = (2p \sin^2\frac{\alpha}{2}+2\sqrt{1-p}\sin \alpha).
\end{eqnarray} 
Since $E_{12}^\prime(\rho_{1})-E_{12}^\prime(\rho_{3})\geq  0$ in the full range  $0\leq p\leq 1$, $E_{12}^\prime(\rho_{1})\geq E_{12}^\prime(\rho_{3})$. The proof of $E_{12}^\prime(\rho_{12})\geq E_{12}^\prime(\rho_{123})$ also follows from the fact that $E_{12}^\prime(\rho_{12})- E_{12}^\prime(\rho_{123})>0$ for all values of $p$, which can be shown by using the expressions of $E_{12}^\prime(\rho_{12})$ and $E_{12}^\prime(\rho_{123}$ as given in Eqs.~(\ref{eq:adneg_1}) and (\ref{eq:adneg_5}). \hfill $\blacksquare$

\bibliography{le_draft_v1}{}
\bibliographystyle{apsrev4-1}

\end{document}